\documentclass[
reprint,
amsmath,
amssymb,
 aps,nofootinbib]{revtex4-2}

\usepackage{graphicx}
\usepackage{dcolumn}
\usepackage{bm}
\usepackage{hyperref}

\usepackage{soul}
\usepackage{braket}
\usepackage[dvipsnames]{xcolor}
\usepackage{siunitx} 

\begin{document}

\title{The implications
of stochastic gas torques for asymmetric binaries 
in the LISA band}

\author{Lorenzo Copparoni}
\email[]{lcopparo@sissa.it}
\affiliation{SISSA - Scuola Internazionale Superiore di Studi Avanzati, Via Bonomea 265, 34136 Trieste, Italy and INFN Sezione di Trieste}
\affiliation{IFPU - Institute for Fundamental Physics of the Universe, Via Beirut 2, 34014 Trieste, Italy}
\author{Lorenzo Speri}
\affiliation{European Space Agency (ESA), European Space Research and Technology Centre (ESTEC), Keplerlaan 1, 2201 AZ Noordwijk, the Netherlands}
\affiliation{Max Planck Institute for Gravitational Physics (Albert Einstein Institute) Am Mühlenberg 1, D-14476 Potsdam, Germany}
\author{Laura Sberna}
\affiliation{School of Mathematical Sciences, University of Nottingham, University Park, Nottingham NG7 2RD, United Kingdom }
\author{Andrea Derdzinski}
\affiliation{Department of Life and Physical Sciences, Fisk University, 1000 17th Avenue N., Nashville, TN 37208, USA}
\affiliation{Department of Physics \& Astronomy, Vanderbilt University, 2301 Vanderbilt Place, Nashville, TN 37235, USA}

\author{Enrico Barausse}
\affiliation{SISSA - Scuola Internazionale Superiore di Studi Avanzati, Via Bonomea 265, 34136 Trieste, Italy and INFN Sezione di Trieste}
\affiliation{IFPU - Institute for Fundamental Physics of the Universe, Via Beirut 2, 34014 Trieste, Italy}

\begin{abstract}
Gravitational waves from asymmetric mass-ratio black-hole binaries carry unique information about their astrophysical environment.  For instance,  
the Laser Interferometer Space Antenna (LISA)  could potentially measure the amplitude and slope of gas torques in binaries embedded in the accretion disks of Active Galactic Nuclei, helping differentiate competing accretion disk models. However, this relies on simplified analytic models, which do not account for the stochastic variability of torques seen in hydrodynamic simulations. In this work, we use hydrodynamic simulations to create gravitational waveforms for extreme and intermediate mass-ratio inspirals in the LISA band. We then analyze these simulated waveforms using simpler templates that assume analytic torques, without stochastic time variability.
By performing realistic Bayesian parameter estimation, we find no bias at 90\% confidence in the binary parameters; however, estimates of accretion disk parameters, such as torque amplitude and slope, may be biased.
Typically, the posterior distribution is centered around the average value of the torques, but when stochastic variability is large, the posterior can indicate no torques, even though they are present in the simulation. Our results suggest that while simplified analytic torque models work well for estimating binary parameters, caution is needed when using them to infer properties of the accretion disk.
This work moves towards a more realistic assessment of one of the LISA science objectives, i.e., probing the properties of the astrophysical environments of black holes.
\end{abstract}

\maketitle
\section{Introduction}
LIGO and Virgo have detected over a hundred compact object binaries to date~\cite{PhysRevX.13.041039,gwtc2:1,gwtc2,gwtc1,PhysRevLett.116.061102,GW170817}, revealing their properties~\cite{KAGRA:2021duu,Abbott_2019,Abbott_2021} and formation channels~\cite{Fchannel:GW150914,Fchannel:GW190521}, and enabling tests of gravity in strong-field regimes~\cite{theligoscientificcollaboration2021testsgeneralrelativitygwtc3,TestGr:GW150914,TestGR:GW170817,TestGR:GWTC1,TestGR:GWTC2}. While most binaries consist of black holes of likely stellar origin, but the most massive system, GW190521~\cite{PhysRevLett.125.101102}, features individual masses (approx.~$85~M_\odot$ and $66~M_\odot$) in the pair instability mass gap~\cite{woosleyPulsationalPairinstabilitySupernovae2017,Farmer_2019}. This suggests an unusual origin, possibly in the disk of an Active Galactic Nucleus (AGN) through repeated mergers~\cite{Tagawa_2020}.
If binaries like GW190521 formed in an AGN disk, interactions with the surrounding gas and the central massive black hole could leave characteristic features in its low-frequency gravitational wave signal. These environmental effects -- including accretion, dynamical friction, peculiar acceleration/Doppler modulation, Shapiro time delay, and lensing~\cite{Toubiana:2020drf,Sberna:2022qbn,caputoGravitationalwaveDetectionParameter2020} -- would be detectable by space-based detectors like the Laser Interferometer Space Antenna (LISA)~\cite{Colpi:2024xhw}. Indeed, LISA will be the perfect instrument to detect the interaction of black hole binaries with their astrophysical environment. This has been highlighted in several works~\cite{Barausse:2007dy,PhysRevD.84.024032,yunesImprintAccretionDiskInduced2011,barausseCanEnvironmentalEffects2014,berry2019uniquepotentialextrememassratio,caputoGravitationalwaveDetectionParameter2020,toubianaDetectableEnvironmentalEffects2021,sbernaObservingGW190521likeBinary2022,speriProbingAccretionPhysics2023}, and is due to the large number of orbital cycles that LISA will observe, which allows environmental effects to accumulate secularly up to being detectable.

Environmental effects are typically expected to be most detectable in asymmetric binaries\footnote{Gas interactions might also be detectable in GWs from equal mass  black hole binary systems, when the components are in the intermediate mass range \cite{Garg2024}.}, consisting of a 
 massive black hole (with mass $M_1$ in the range $10^5$--$10^7 M_\odot$) and a smaller black hole with mass $M_2\gtrsim 10 M_\odot$ \cite{babakScienceSpacebasedInterferometer2017}.  Asymmetric binaries are conventionally classified as extreme mass-ratio inspirals (EMRIs) or intermediate mass-ratio inspirals (IMRIs)
 according to the mass ratio $q\equiv M_2/M_1\leq 1$, with IMRIs ($q\gtrsim 10^{-3}$) traditionally believed to be less common than EMRIs.
 Upper limits for IMRI rates in LISA are estimated around tens per year and are subject to large uncertainties \cite{LISA:2022yao}.
 This is because there is currently no solid evidence from electromagnetic observations for black holes with intermediate masses ($100~M_{\odot}$ to $10^5~M_{\odot}$) \cite{Greene:2019vlv}.
 However, the discovery of GW190521, with its large individual and remnant masses, and the observation of quasi periodic emission/outflows in AGNs~\cite{10.1093/mnras/stae1599}, has made the detection of IMRIs more likely.
In particular, the observations of X-ray flare ASASSAN-20qc, followed by a quasi periodic oscillation, are well compatible with the motion of an intermediate mass black hole perturbing the AGN~\cite{doi:10.1126/sciadv.adj8898}.

 The dominant environmental interaction affecting the evolution of EMRIs/IMRIs is the interaction of the secondary object with the primary's accretion disk. Indeed, a significant fraction ($1-10\%$) of massive black holes in the local universe shine as AGNs as a result of radiatively efficient accretion~\cite{Dittmann2020}. Hence, one would naively expect at least a comparable fraction of the EMRI/IMRI population to be residing in AGN disks~\cite{Levin:2007,chakrabartiGravitationalWaveEmission1996}.
Moreover, several works have also highlighted disk-driven EMRI formation channels, beyond the standard loss cone mechanism~\citep{shapiroStarClustersContaining1978,bar-orSTEADYSTATERELATIVISTICStelLAR2016}, such as accretion disk capture~\citep{Tagawa_2020,Pan:2021oob, Pan:2021ksp} or the birth of stars from accretion disk instabilities~\citep{Levin2007,GoodmanTan2004,derdzinskiInsituExtremeMass2023}.

 In recent years, a number of works have investigated the scientific potential of EMRIs in AGN accretion disks, concluding that LISA will be able to detect gas-driven migration and extract disk properties~\cite{Barausse:2007dy,PhysRevD.84.024032,yunesImprintAccretionDiskInduced2011,barausseCanEnvironmentalEffects2014,Zwick2022,speriProbingAccretionPhysics2023,Duque:2024mfw,Kejriwal:2023djc,Khalvati:2024tzz}. However, all these works rely on simple analytic models for the gas torques driving the secondary's migration, neglecting, among other features, the stochastic time variability of the torques observed in more realistic hydrodynamic simulations~\cite{derdzinskiProbingGasDisc2019,duffellMigrationGapOpeningPlanets2014,Wu:2023qeh}.

In this paper, we fill this gap, and use for the first time numerically simulated disk torques, including their stochastic fluctuations, to generate more realistic EMRI/IMRI trajectories and GW signals. Our stochastic disk torques are extracted from the hydrodynamic simulations of Ref.~\cite{derdzinskiEvolutionGasDisc2021}.
We then investigate whether adopting simple analytic models for the torques, which ignore the fluctuations, can bias the recovery of the EMRI/IMRI parameters and the properties of the accretion disk. 
We do so using a Bayesian parameter estimation pipeline meant for the LISA mission.

We find that no bias at 90\% confidence level affect the binary parameters, but that the recovered amplitude and slope for the torques can be different from their ``true'' values due to the stochastic time variability. While in some cases our parameter estimation produces posteriors centered on the torque amplitude averaged over the time variability, for one specific case with large stochastic variation our pipeline finds no evidence of the torques in the signal, missing the environmental interaction altogether.
We conclude that caution should be used when employing simple analytic torque models to extract information on accretion disks from EMRI/IMRI data, as we find significant biases in the estimation of the disk surface density.

 This paper is organized as follows. In section~\ref{sec:plan_mig} we provide a quick review of Type I and Type II planetary-like migration. In section~\ref{sec:num_sig} we describe the methods used to generate the signals undergoing gap-opening migration and the set-up used for the analysis.
 Finally, in section~\ref{sec:results} we discuss our results, while in section~\ref{sec:astro} we  provide a short discussion of possible biases that could arise when trying to measure the astrophysical environment's properties.
Throughout this work, we make use of geometric units  $G = c = 1$, unless otherwise stated.

\section{Disk driven Migration}\label{sec:plan_mig}
The gravitational field generated by a  compact object  moving in a gaseous environment produces spiral density perturbations through Lindblad and corotation resonances~\cite{Goldreich1979}, resulting
in an exchange of angular momentum between the satellite and the disk~\cite{goldreichDisksatelliteInteractions1980}.
This effect produces an inward, or outward,  migration of the secondary object, and has been extensively studied in the context of planetary formation~\cite{baruteauRecentDevelopmentsPlanet2013,nelsonPlanetaryMigrationProtoplanetary2018}.

The case in which the density perturbations do not strongly modify the disk structure is commonly referred to in the literature as Type I migration.
This regime is relevant  when the mass of the satellite is much smaller than the mass of the primary surrounded by the disk.
Under these assumptions, one can neglect any effect due to the radial motion of the satellite when studying the torque exchange between the disk and the secondary~\cite{baruteauPlanetDiscInteractionsEarly2014}.
The torque has been modelled at linear order in 2 and 3-dimensional isothermal disks in Refs.~\cite{goldreichDisksatelliteInteractions1980,tanakaThreeDimensionalInteraction2004}. In the case of a satellite on a circular orbit migrating in an isothermal disk, the torque is given by~\cite{tanakaThreeDimensionalInteraction2002}
\begin{equation}
	\dot{L}_I = \Sigma(r)r^4 \Omega^2 q^2 \mathcal{M}^2
	\label{eq:Tanaka_torque}
\end{equation}
where $\Omega$ is the angular velocity of the satellite,  $\mathcal{M}$ is the Mach number of the disk, and $\Sigma$ is the unperturbed surface density profile. For realistic disks, we expect additional coefficients multiplying the linear torque, depending on the temperature and density gradient~\cite{PhysRevD.90.084025,Tanaka_2024}. In this work, we ignore those additional terms.
This expression reproduces  the results of numerical hydrodynamical simulations~\cite{korycanskyNumericalCalculationsLinear1993}.

When considering the migration of more massive objects, it is possible for the torque exerted by the satellite on the disk to become larger than the viscous torque, which is responsible for disk spreading~\cite[pag.~749-835]{Levy1993ProtostarsAP}.
The density wave perturbations launched by the satellite are then able to carve a low density annular region, also known as gap, around the  orbit~\cite{Lin1986}.
In this case, the system is said to be undergoing Type II (or gap opening) migration.

It should be noted that  gap formation alone does not automatically imply that the system undergoes  Type II migration.
It is possible for the gap to  close too quickly to provide a significant contribution to the satellite dynamics. 
The conditions separating the two regimes depend non-trivially on the disk parameters and the mass ratio of the system~\cite{Malik2015}. In more detail, the following conditions are necessary for a system to undergo type II migration:
\begin{itemize}
	\item the secondary perturbation pushing away the gas cannot be compensated by the gravitational attraction and the viscosity of the disk~\cite{cridaWidthShapeGaps2006};
	\item the gap formation time must be smaller than the time necessary for the secondary to cross its horseshoe radius~\cite{Ward1989};
	\item the gap formation time must be smaller than the viscous diffusion timescale~\cite{gargImprintGasGravitational2022a}.
\end{itemize}
While the first condition concerns the gap formation itself, the last two ensure that the gap is preserved during the migration and is not refilled by viscous diffusion.

There have been numerous numerical studies on planetary evolution in the gap formation regime~\cite{nelsonMigrationGrowthProtoplanets2000,Duffell2013}.
Various empirical approximations describing the torque exchange have also been proposed, such as the one in Ref.~\cite{kanagawaRadialMigrationGapopening2018}, 
\begin{equation}
	\dot{L}_{II} = \frac{\dot{L}_{I}}{1+0.04q^2\mathcal{M}^5\alpha^{-1}}
	\label{TypeII_Linear}
\end{equation}
where $\alpha$ is the viscosity of the disk.
These empirical formulae, however, fail to reproduce the full non-linear behavior observed in numerical simulations,
particularly in the parameter space of high-viscosity, geometrically thin disks.
The full non-linear contribution can indeed lead to a sign change in the net exchanged torque~\citep{derdzinskiProbingGasDisc2019,duffellMigrationGapOpeningPlanets2014}.
While a fully satisfactory analytic description has not been put forward yet, a common feature observed in hydrodynamical simulations is a dampening of the mean torque exerted on the disk when compared with the Type I equation~\eqref{eq:Tanaka_torque}.

Moreover, in realistic astrophysical situations, 
stochastic features are generically expected in 
the evolution of  proto-planetary systems and in the motion of stellar mass and intermediate mass black holes in AGN disks. 
Fluctuations arise both from the influence of the secondary and from the turbulent nature of the disk.
Turbulence becomes even more important for  asymmetric mass ratio systems undergoing Type I migration~\cite{nelsonOrbitalEvolutionLow2005,Baruteau_2010,WuChenLin2024}, 
where it can give rise to \emph{chaotic} migration.
As the mass ratio grows and the migration enters the gap-opening regime, the chaotic turbulence contribution becomes instead less significant~\cite{Wu:2023qeh,Aoyama_2023}. 

\section{Injected signals and templates}
\label{sec:num_sig}
In this section, we describe the setup used to generate the injection, and the parameter estimation framework.
First, in~\ref{sec:inj} we briefly review the simulations of Ref.~\cite{derdzinskiEvolutionGasDisc2021}, and describe how we implement the numerical torque in the injected waveforms.
Then, in~\ref{sec:data_analysis} we describe the template family used in the parameter estimation.

\subsection{Injection}
\label{sec:inj}
Our review of Type II migration suggests that, to obtain a  realistic estimate of the disk torques, it is necessary to perform fully numerical hydrodynamical simulations.
In this work, we use the numerical results obtained by~\citeauthor{derdzinskiEvolutionGasDisc2021} in~\cite{derdzinskiEvolutionGasDisc2021}.
In the latter, they considered a thin, isothermal Shakura-Sunyaev accretion disk, with constant Mach number $\mathcal{M} = (H/r)^{-1}$ and constant $\alpha$-law prescription for the viscosity.
Imposing these conditions in the equations describing the thin disk structure in the Newtonian limit~\cite{frankAccretionPowerAstrophysics2002},  it follows that the surface density profile is $ \Sigma \propto r^{-1/2}$.

The secondary was modeled as a point mass with a smoothed gravitational potential, which inspirals
on quasicircular equatorial orbits under gravitational radiation reaction (computed using the quadrupole formula). 
The effect of the environmental torque exerted on the secondary was neglected in the trajectory evolution.
The simulations are scale invariant with respect to both the total mass of the system and the surface density of the accretion disk. 
Because the simulations are two-dimensional, they require that the smoothing of the secondary potential is of order a scale height of the disk in order to properly model vertically-integrated forces. Specifically, it is set to be one half of the disk scale height, as adopted in other studies \cite{tanakaThreeDimensionalInteraction2002,Masset2002}. For the case of a secondary of $10^3 M_\odot$, this corresponds to a scale of $\sim 275$ Schwarzschild radii below which the gas morphology is not resolved. The variability of the torques measured in this study is due to the gas flow outside of this region, and persists with higher resolution tests \cite{derdzinskiProbingGasDisc2019}.

Here, we instead consider quasicircular and equatorial prograde orbits around a {\it spinning} primary,  evolving adiabatically under the effect of gravitational wave (GW) fluxes~\cite{Teukolsky:1973ha,PhysRevD.90.084025,drascoGravitationalWaveSnapshots2006} down to the Kerr ISCO~\cite{bardeenRotatingBlackHoles1972}. 
We do not consider the inspiral-to-plunge and ringdown GW emission, as they contain negligible power for very asymmetric mass ratio binaries.
We rescale the simulations, which were performed in a scale-free Newtonian setting, so that the disk's inner edge 
matches the Kerr ISCO.
We also interpolate the simulation torque in  orbital separation $r$  
rather than time.
This choice is mainly dictated by the fact that we are using adiabatic GW fluxes, while in the numerical simulations the inspiral was modeled using only the GW quadrupole flux.
Finally, we add the environmental torque to the GW fluxes and evolve the trajectory with the \texttt{FastEMRIWaveform} (FEW) package~\citep{Chua:2020stf,katzFastEMRIWaveformsNewTools2021,Speri:2023jte}.

\begin{figure*}
    \centering
    \includegraphics[width = 1.0\textwidth]{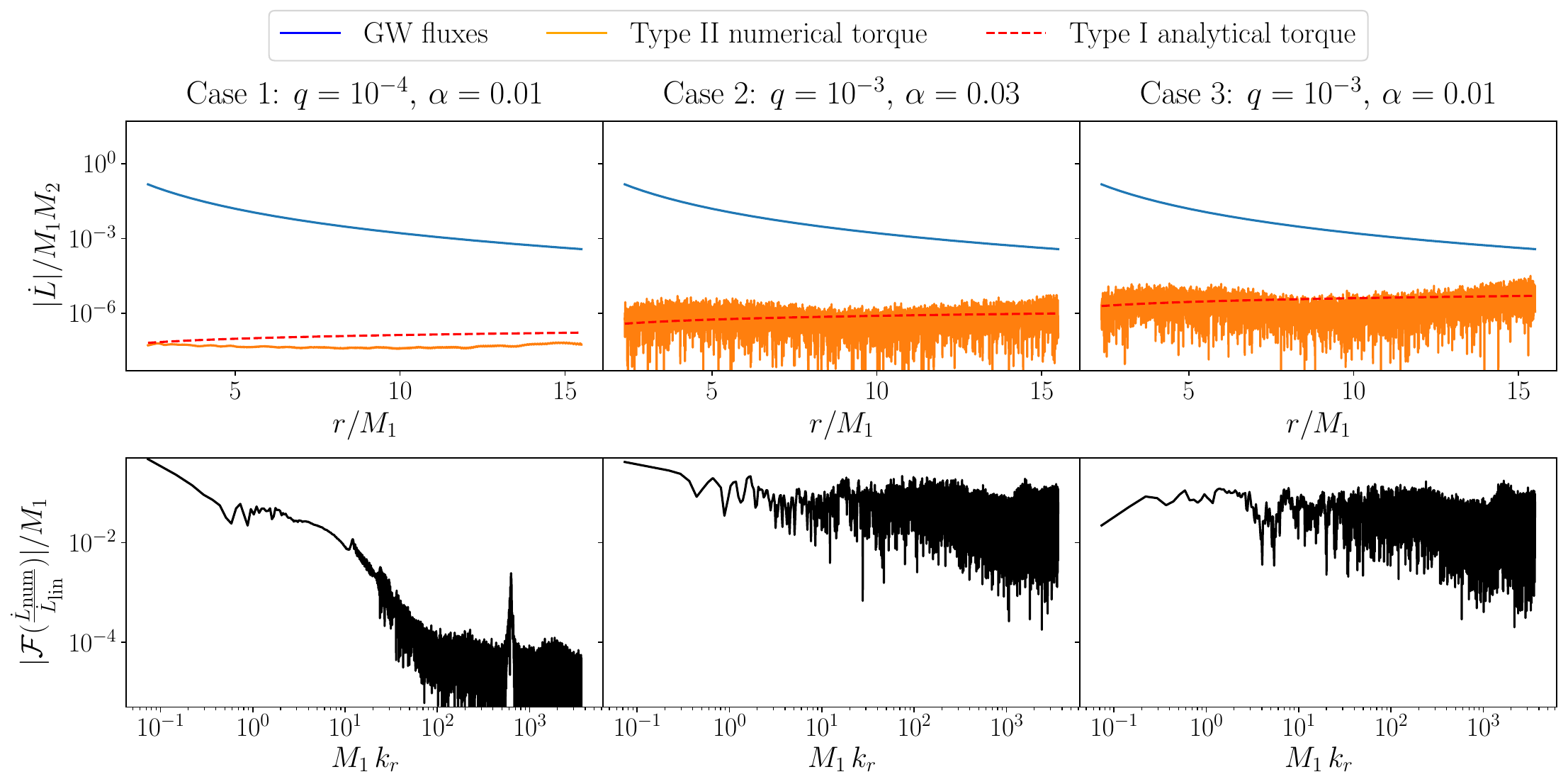}
    \caption{
    Impact of stochastic torques on the binary evolution. 
    In the top panels we report the different components responsible for the binary evolution: the torque due to the GW fluxes (blue) and the disk  torque (orange).
    The red dotted lines represent the analytic estimate of the linear (Type I) disk torque, Eq.~\eqref{eq:Tanaka_torque}.
    Note that, in all cases studied here, the environmental effect is subleading compared to  GW emission.
    In the bottom panels, we plot the absolute value of the  Fourier transform of the ratio between the numerical  and  analytic disk torques.   }
    \label{fig:torque}
\end{figure*}

We consider three cases associated with the simulations of~\cite{derdzinskiEvolutionGasDisc2021}, which vary by mass ratio and disk viscosity $\alpha$:
\begin{itemize}
	\item \textbf{Case 1:} $q=10^{-4},\,\alpha = 0.01$,
	\item \textbf{Case 2:} $q=10^{-3},\,\alpha = 0.03$,
	\item \textbf{Case 3:} $q=10^{-3},\,\alpha = 0.01$\,.
\end{itemize}
In all cases, the Mach number is $\mathcal{M} = 20$. 
The complete list of our systems' parameters is reported in Table~\ref{tab:models}. In all cases, we select the initial separation $r_0$ so that the secondary plunges within the observation time.
We choose a random sky position and spin orientation, while we choose the distance to set the signal-to-noise ratio (SNR) of the signals to the values given in the next Section.

\begin{table} 
	\centering
	\begin{tabular}{cccc}
		\hline
              & Case 1     & Case 2    & Case 3     \\
                \hline
$M_1/M_\odot$           & $2.2\times 10^6$ & $10^6$ & $2\times 10^7$ \\
$q$ 	      & $10^{-4}$  & $10^{-3}$ & $10^{-3}$  \\
$a/M_1$ 	      & $0.9$      & $0.9$     & $0.9$      \\
$r_0/M_1$      & $15.10$    & $15.487$  & $15.487$   \\ 
$\Phi_S$      & $1.43$     & $2.30$    & $0.39$     \\
$\theta_S$    & $1.88$     & $2.85$    & $0.54$     \\
$\Phi_K$      & $1.59$     & $1.84$    & $5.67$    \\
$\theta_K$    & $3.059$    & $0.95$    & $0.04$     \\
$\Phi_0$      & $5.45$     & $3.59$    & $2.69$     \\
$\Sigma_0~(\si{\kilo\gram\per \metre^2})$ & $1.40\times 10^7$ & $1.8\times 10^8 $  & $ 4.6\times10^7$\\
$\alpha$      & $ 0.01$    & $0.03$    & $0.01$     \\
$T\textsubscript{obs}$ (\mbox{yr}) & 4.0 & 0.2 & 4.0\\
$\mathcal{M}$ & $20.0$     & $20.0$    & $20.0$     \\
		\hline
	\end{tabular}
	\caption{Parameters for the three different systems that we consider. $\Sigma_0$ is the surface density of the disk at $r=10M$.  
	\label{tab:models}}
	\end{table}
To generate the injected waveforms, 
we use the augmented analytic kludge (AAK) algorithm~\cite{PhysRevD.96.044005} implemented in FEW, together with an adiabatic trajectory modified for the purpose of this work as in \cite{speriProbingAccretionPhysics2023}. 
The AAK algorithm is a refinement of the analytic kludge model~\cite{barackLISACaptureSources2004}, which approximates the secondary orbits as Keplerian ellipses, and calculates the waveform using the Peters-Mathews mode sum approximation~\cite{peters_gravitational_1963}. The AAK algorithm improves the accuracy by using the Kerr orbital frequencies in place of their Keplerian approximations.

Following Ref.~\cite{derdzinskiEvolutionGasDisc2021}, we select $\Sigma_0$ such that the environmental contribution is subdominant relative to the GW fluxes.
Note that this results in realistic values for the density, $\Sigma_0\sim 10^7-10^8 \mbox {kg/m}^2$~\cite{frankAccretionPowerAstrophysics2002}.
In Fig.~\ref{fig:torque}, in the top panels, we compare the contributions to the torque arising from the GW fluxes and the environment.  
As can be seen, the analytic expression for  Type I torques [Eq.~\eqref{eq:Tanaka_torque}], represented by a red dashed line, cannot capture the stochastic behavior of  the numerical torque (in orange), which is dominant for $q=10^{-3}$. 

Also shown in the bottom panels of
Fig.~\ref{fig:torque}, is
the absolute value of the Fourier transform of the ratio between the numerical and the analytic disk torques. 
As can be seen, the stochastic torques (middle and left panels) have a rather flat spectrum. The torque for case 1, which shows no stochastic features, has a peak at
wavenumber $M k_r\sim 10^3$, resulting from small
oscillations barely visible in the top panel. The low frequency bump in the left panel is instead due to the torque's average slow evolution across the plotted radial range. 
The lower panel of Fig.~\ref{fig:torque} also shows that the smaller viscosity in case 3 suppresses the average torque by almost two orders of magnitude compared to case 2, see the low-frequency end, despite similarities in their stochastic fluctuations. 

Note that for cases 1 and 2, we choose masses  $M_1=2.2\times 10^6~M_\odot$ and $M_1=10^6~M_\odot$ respectively, while for case 3 we consider a more massive primary  $M_1=2\times10^7~M_\odot$. This latter choice allows for enhancing the 
 stochastic torque, which depends linearly on $M_1\Sigma_0$~\cite{derdzinskiEvolutionGasDisc2021}, while 
 keeping the injected signal marginally within the LISA band.
This enhancement of the stochastic torque is not necessarily unrealistic. Indeed, Ref.~\cite{derdzinskiEvolutionGasDisc2021} 
can only simulate disks with Mach numbers $\mathcal{M}\simeq30$, corresponding to thicker geometries than realistic AGN disks, for which $\mathcal{M}$ can reach $\sim 100$. If realistic disk aspect ratios could be simulated, they could possibly present more pronounced stochastic torque components, 
if the trend in the simulation data continues to the regime of highly supersonic flows. We note that the simulations discussed only consider laminar disks, finding that torque variability is due to the instabilities in the gas flow around the secondary. Realistic AGN disks are highly magnetized and turbulent, and this may lead to an additional source of torque stochasticity \cite{Baruteau_2010,WuChenLin2024,zwickDirtyWaveformsMultiband2022}. However, more research is needed to investigate this regime for the AGN IMRI case, in which there is likely an interplay between the nature of the turbulence, the characteristics of the disk and the BH mass.

\subsection{Template and Parameter estimation}
\label{sec:data_analysis}
To perform parameter estimation, we build our templates using the AAK algorithm and Kerr circular equatorial orbits,
supplemented with an environmental torque. 
For the latter, 
we adopt a simple parametric model inspired by the analytic Type I expression~\eqref{eq:Tanaka_torque} \cite{speriProbingAccretionPhysics2023}, i.e.
 we supplement the GW flux with a power-law term given by
\begin{equation}
    \dot{L} = \dot{L}\textsubscript{GW}  \, A\bigg(\frac{r}{10M_1}\bigg)^n\,.
    \label{eq:env_template}
    \end{equation}
Note that because the torque is normalized to the GW flux
$\dot{L}\textsubscript{GW}=-32/5\,q(r/M_1)^{-7/2}$,
the Post-Newtonian (PN) order 
at which environmental effects appear is given by $-n$.
For consistency with the reference radius
appearing in \eqref{eq:env_template}, 
in the following we will denote by $\Sigma_0$ the surface density at $r=10M_1$.

To perform Bayesian parameter estimation, we inject the signal in the parallel tempering Monte Carlo Markov chain (MCMC) sampler \texttt{eryn}~\citep{karnesisErynMultipurposeSampler2023}.
We  use 4 temperatures and 16 walkers, and we stop the sampler only if the chain length is at least 50 times the mean autocorrelation time evaluated across chains.
We sample over the intrinsic parameters ($\log M_1/M_\odot$, $\log M_2/M_\odot$, $a/M_1$, $r_0/M_1$,$\Phi_{\phi0}$), with $a$ being the primary spin and $r_0,\,\Phi_{\phi 0}$ the initial separation and phase for $\phi$; the extrinsic parameters ($D_L$, $\cos\theta_S$, $\Phi_S$, $\cos\theta_K$, $\Phi_K$), with $D_L$ being the luminosity distance, $\theta_S$, $\Phi_S$ being the polar and azimuthal sky position in the Solar System barycenter frame, $\theta_K$, $\Phi_K$ being the polar and azimuthal angle describing the orientation of the primary spin; and the power-law parameters ($A,\,n$).
We adopt the standard Gaussian likelihood $\log\mathcal{L} = -1/2\sum_{X \in {\rm (I,\,II)}}\braket{s_X-h_X(\mathbf{\lambda})|s_X - h_X(\mathbf{\lambda}) }$, with  $s$  the data  in the two LISA data channels in the long-wavelength approximation~\cite{cutlerAngularResolutionLISA1998}, and $h(\mathbf{\lambda})$ the GW template with parameters $\mathbf{\lambda}$.
The inner product is defined as~\cite{cutlerGravitationalWavesMerging1994}:
\begin{equation}
    \braket{a|b} = 4 \mathrm{Re}\int_0 ^{+\infty}\mathrm{d}f\, \frac{\tilde{a}^*(f)\tilde{b}(f)}{S_n(f)} \, ,
\end{equation}
where $S_n(f)$ is LISA's noise power spectral density (PSD)~\cite{SciRd}, while the tilde indicates the Fourier transform.

\section{Results}\label{sec:results}
In this section, we review the 
parameter estimation results for the three systems mentioned above. 
For all three systems, we will consider  $n=4$ as the ``expected value'' of the power law slope, as predicted by Eq.~\eqref{eq:Tanaka_torque} for the disk model of the simulations.
To estimate the ``expected value'' of the power-law amplitude, we use the  torque
averaged over the radial range of the simulations, assuming the expected slope $n=4$. This procedure results in a smaller amplitude than
would be predicted by the Type I expression~\eqref{eq:Tanaka_torque}, by a factor of $\sim0.28$, $\sim-0.22$ and  $\sim-0.015$ for cases 1, 2 and 3, respectively.
It was observed in~\cite{derdzinskiEvolutionGasDisc2021} that in the last two cases the net migration is outward rather than inward, even in the last stages of the inspiral.

\begin{figure}
    \centering
    \includegraphics[width =1.0\linewidth]{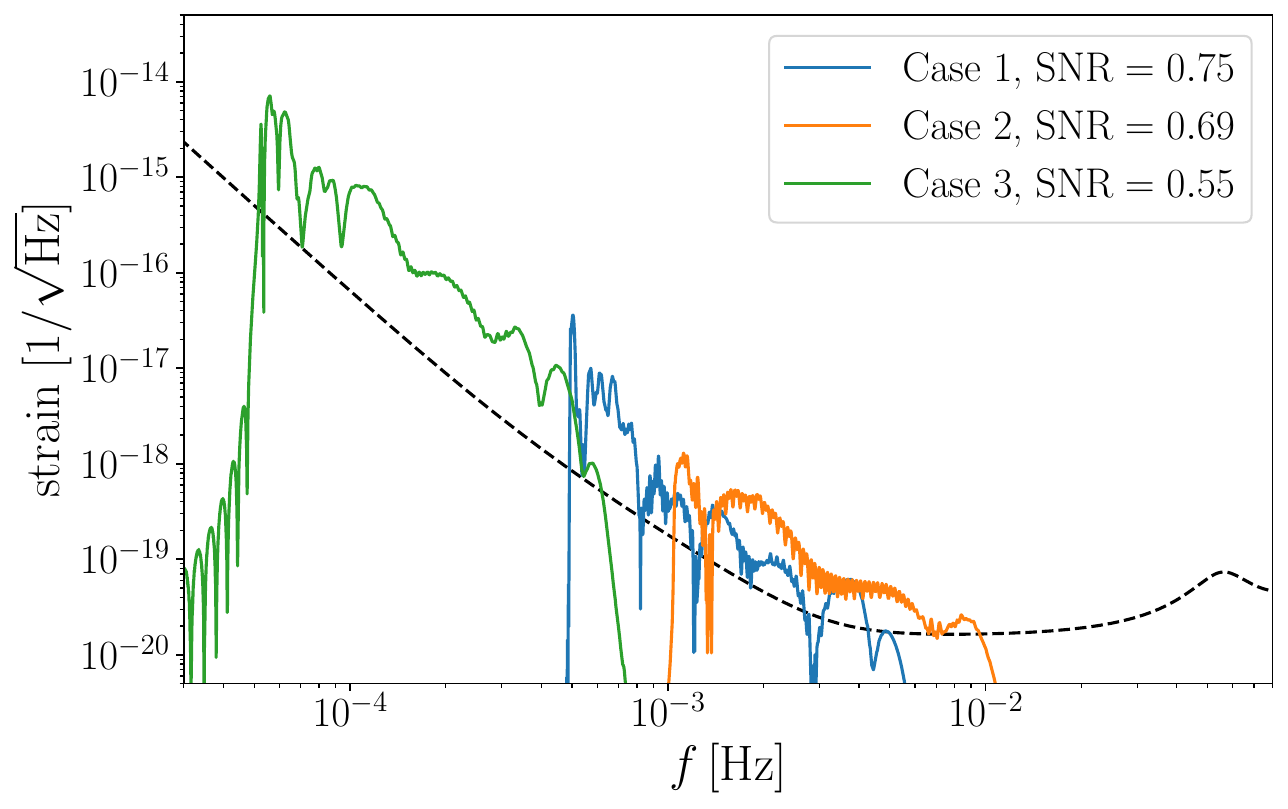}
    \caption{
    Residual signal left by subtracting the best-fit waveform obtained with  analytic-torque templates from a signal produced with simulated torques. The SNR of the residuals is 
    reported in the legend for the three systems we consider, summarised in Table~\ref{tab:models}.}
    \label{fig:unresolved}
\end{figure}

In Fig.~\ref{fig:unresolved} we show the residual signal  left in the data after subtracting the  best-fit template.
In all the cases studied, the residual SNR is negligible ($\lesssim 1$). As such, it should not affect significantly the LISA global fit~\cite{Strub:2024kbe,Littenberg:2023xpl,Katz:2024oqg,Deng:2025wgk}.
\subsection{Case 1 ($q=10^{-4}$)}
First, we consider the system with mass ratio $q=10^{-4}$.
Here, the stochastic features are subdominant relative to the mean contribution, i.e. they can be modeled as small perturbations to the mean torque, which has approximately power-law radial dependence, see Fig.~\ref{fig:torque}.
We set the primary mass to $M_1=2.2\times 10^6~M_\odot$, the disk surface density to $\Sigma_0 = 1.40\times 10^{7}~\si{\kilo\gram\per \meter ^ 2}$, so that we expect the power-law amplitude to be $A = 2\times 10^{-5}$.
Using the parameters indicated in Table~\ref{tab:models} and an observation time of $4~\si{yr}$, we fix the SNR for this event to 100 by setting the luminosity distance to $2.26~\si{\giga pc}$.

We show the posterior distribution of ($\log M_1/M_\odot$, $\log M_2/M_\odot$, $a/M_1$, $A$, $n$) in Fig.~\ref{fig:posterior_q1e4_intr}, while in Fig~\ref{fig:posterior_q1e4} of the Appendix we show the full posterior. The posterior of the vacuum parameters does not exhibit biases at 90\% confidence level.
We also find that the posterior of the power-law parameters is compatible at one $\sigma$ with the expected values of $A$ and $n$.

\begin{figure*}
    \centering
    \includegraphics[scale = 0.45]{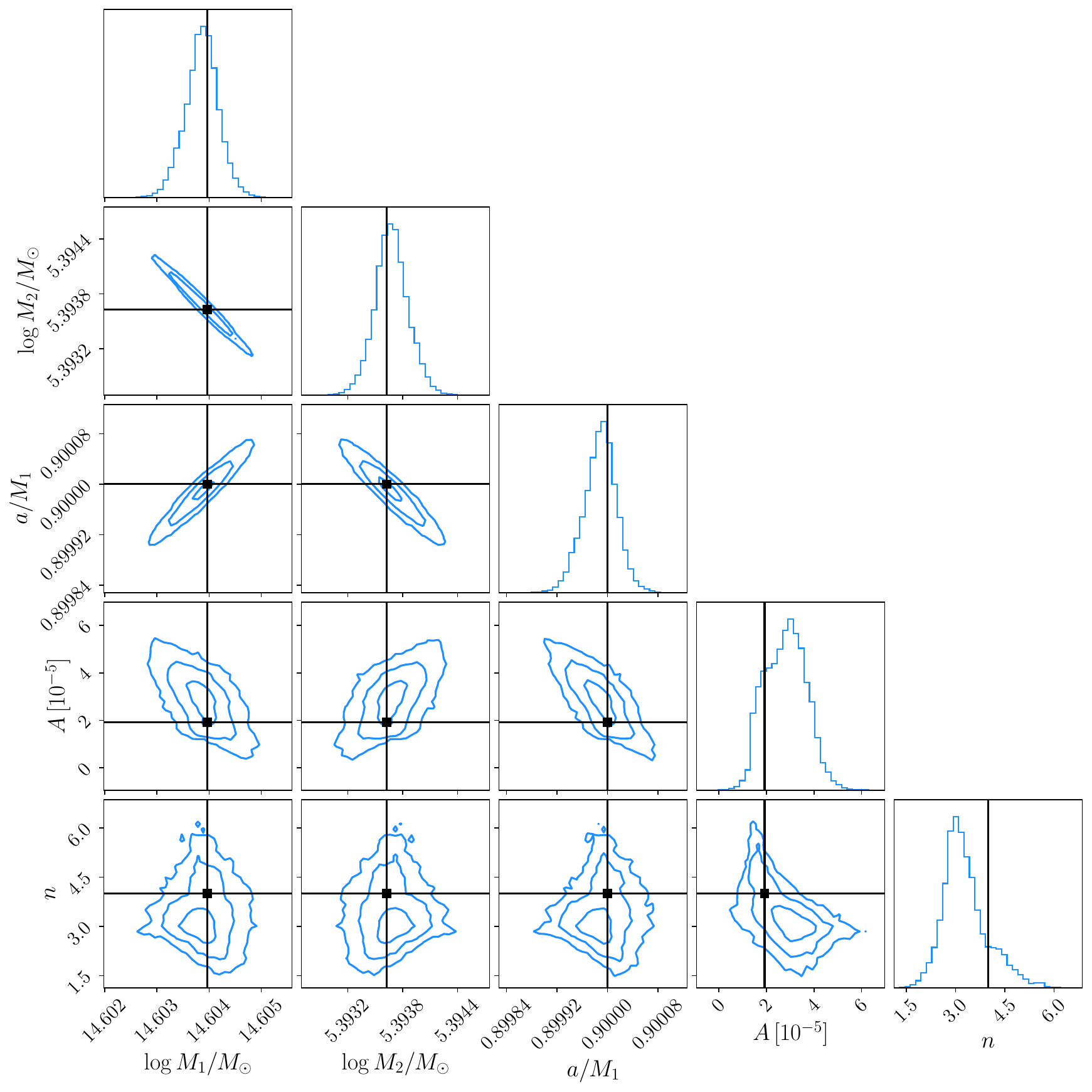}
   \caption{Posteriors of the intrinsic parameters for  case 1 ($q=10^{-4}$). In black we show the true values of the masses and primary spin, and the expected power law amplitude and slope as estimated from the numerical simulations used for the injection~\cite{derdzinskiEvolutionGasDisc2021}.
   We see that using an analytic power law template, one can correctly recover the masses and the spin of the primary. }
    \label{fig:posterior_q1e4_intr}
\end{figure*}

\subsection{Cases 2 and 3 ($q=10^{-3}$)}
In the systems with mass ratio $q=10^{-3}$, the stochastic component dominates over the mean torque, i.e. the environmental contribution is mainly stochastic. Still, we conjecture that it is possible to separate the migration torque into a mean evolution, which follows a power law, and a dominant  noise term, which averages to zero.

We look first at case 2, with primary mass $10^6 M_\odot$ and the remaining intrinsic parameters reported in Table~\ref{tab:models}.
Since we are interested in exploring the impact of stochastic effects on the parameter estimation accuracy, we consider an event of relatively large SNR=200 with an observation time of $0.2~\si{yr}$, which corresponds to a luminosity distance $D_L = 1.375~\si{\giga pc}$.
The duration of this event is significantly shorter than that of typical EMRIs  discussed in the literature.  Indeed, this system 
has larger $q$ and thus larger GW flux (which is proportional to $q$),
which results in a shorter signal duration. 
We choose a disk surface density $\Sigma_0 = 1.8\times 10^{8}~\si{\kilo\gram\per \meter^2}$, which leads to an expected power-law amplitude $A = -9.0\times 10^{-5}$.

\begin{figure*}
    \centering
    \includegraphics[scale =0.45]{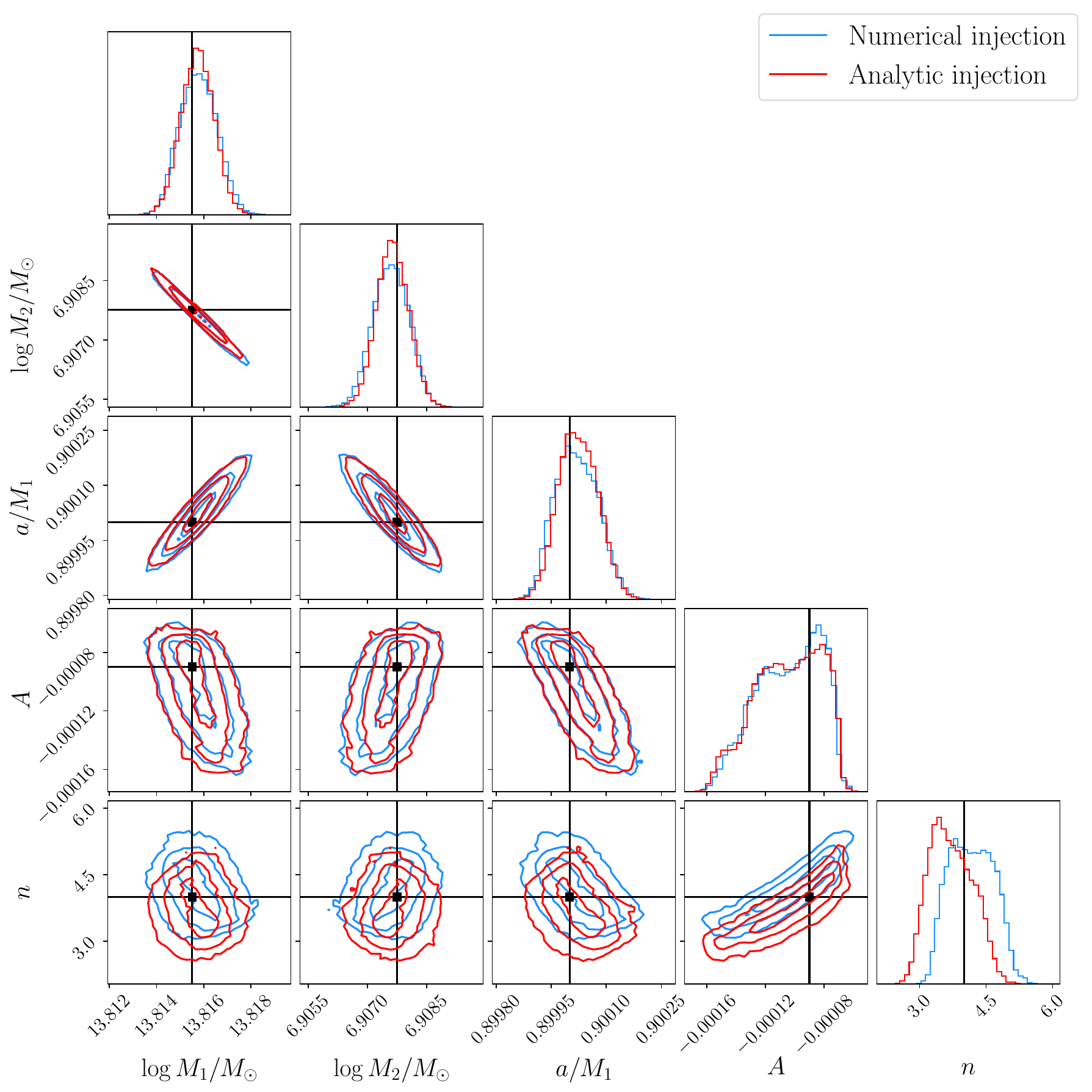}
    \caption{We compare the posteriors of two injections: one in which the IMRI experiences the full stochastic torque of case 2, and one in which the environmental torque is given by the power-law of Eq.~\eqref{eq:env_template}, with amplitude matching the (averaged) numerical torque and slope $n=4$.
    The posteriors have a similar size and are compatible at 90\% confidence level.}
    \label{fig:posterior_comparison}
\end{figure*}

Like in case 1, the posterior of all vacuum parameters, shown in Fig.~\ref{fig:posterior_q1e3_200} of App.~\ref{app:posteriors}, does not exhibit biases.
The parameters of the power-law environmental model are also very well compatible with the expected vales, see Fig.~\ref{fig:posterior_comparison}.

As a further check, we consider a system with the same parameters, but with environmental torque exactly given by the Type I equation~\eqref{eq:Tanaka_torque}, rather than by the numerical simulation.
For consistency, we choose the parameters to match the average torque of the numerical simulation. The posteriors obtained from the numerical and analytic injections are similar, i.e. we can estimate the parameters with a similar precision.
We show this in Fig.~\ref{fig:posterior_comparison}.
We therefore conclude that for a signal undergoing Type II migration, stochastic features in the orbital evolution are likely unimportant, i.e.
one cannot distinguish an analytic injection from a stochastic one.

For case 3,  the average of the numerical torque is two orders of magnitude smaller than  expected from the analytic expression, see Fig.~\ref{fig:torque}.
Because of this strong suppression of the mean torque, the environmental effect will only be detectable by LISA for very dense disks, or for signals with large SNRs and long inspirals in band.
To avoid unrealistically dense disks, we consider a very massive binary with $M_1= 2\times 10^7~M_\odot$, observed for $4~\si{yr}$; by requiring an SNR of 100, we set the event at $D_L = 2.13~\si{\giga pc} $.
We choose an average amplitude of the environmental torque of $A=-2.8\times 10 ^{-5}$ by setting $\Sigma_0 = 4.61\times 10^7 ~\si{\kilo\gram \per m ^ 2}$; see Table~\ref{tab:models} for the remaining parameters. 

\begin{figure*}
    \centering
    \includegraphics[scale =0.45]{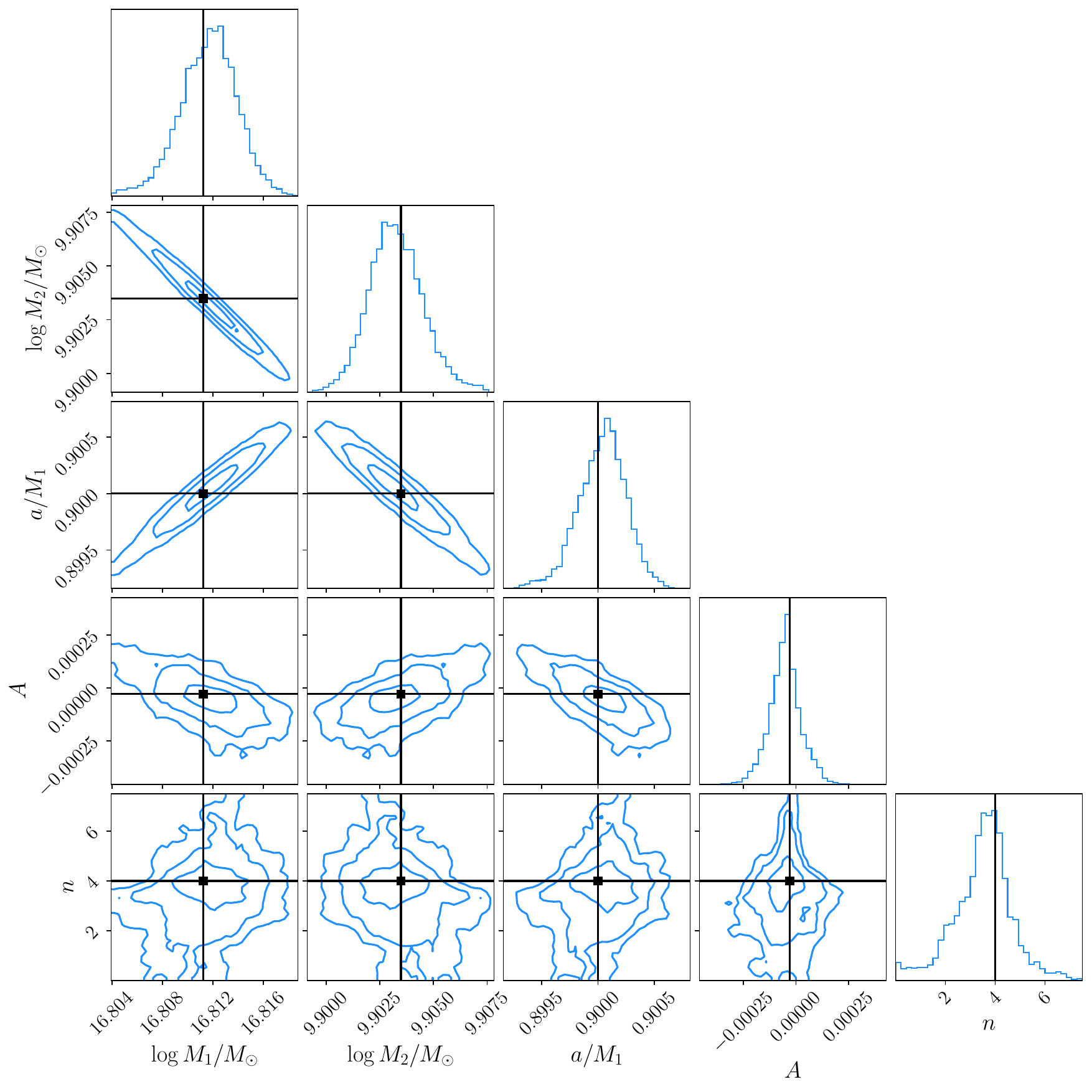}
    \caption{Posterior of the intrinsic parameters for the case 3 ($q=10^{-3}$) injection. We show the injection values (black) of the masses and the primary spin, and the power law amplitude and slope estimated from the hydrodynamical simulations used in the injection~\cite{derdzinskiEvolutionGasDisc2021}.
   Although the vacuum parameters priors are well reconstructed in this case, the stochastic features in the environmental torque wash away the power-law contribution.
   }
    \label{fig:posterior_q1e3_a01_intr}
\end{figure*}

As for the other systems, no significant bias can be observed in the posterior of the vacuum parameters, shown in Fig.~\ref{fig:posterior_q1e3_a01_intr}, while the full posterior is shown in~\ref{fig:posterior_q1e3_a01} of the Appendix.
Even with our choice of a very  massive 
primary and dense disk, the best-fit torque amplitude is smaller than the error inferred  by the parameter estimation.
Therefore, the  posteriors are compatible with vacuum (i.e, the amplitude $A$ is compatible with zero).

Although denser disks would give larger environmental effects, 
one must be cautious because above a certain density threshold the torques may dominate over the gravitational fluxes in the early inspiral, thus violating one of the assumptions of the simulations of~\cite{derdzinskiEvolutionGasDisc2021}.

\section{Astrophysical inference}\label{sec:astro}
In the previous section, we measured deviations from a vacuum signal in a model-independent way. The parameters of our agnostic model can then be mapped to a specific disk model, to  estimate its properties. This inference of the disk
parameters, however, can itself be biased,
as a result of two competing effects: \textit{(i)} the biased estimation of the torques amplitude and slope; and/or \textit{(ii)} the use of the wrong disk model in the inference.

To isolate these two contributions, in this
section we will attempt to reconstruct the disk properties in two ways: \textit{(i)} from the posteriors for $A$ and $n$
obtained for an injection featuring stochastic torques, but utilizing a disk model with constant Mach number (as assumed in the simulations of Ref.~\cite{derdzinskiEvolutionGasDisc2021});
\textit{(ii)} from the posteriors for $A$ and $n$
obtained for an injection featuring analytic torques with no stochastic component, but utilizing a disk model with constant opacity (as assumed e.g. in Ref.~\cite{frankAccretionPowerAstrophysics2002}).
We stress that both disk models are solutions of the (Newtonian) hydrodynamics equations, and are therefore both viable theoretical models to use in the reconstruction of the disk's properties.

The slope of the power-law, $n$, is related to the disk model, as well as to the nature of the environmental effects that modify the inspiral trajectory \cite{PhysRevD.84.024032}. In all the cases that we studied, this parameter does not show biases larger than $2\sigma$, implying that the disk model could be, in principle, accurately constrained.
The power-law amplitude $A$ depends instead, at linear order, on a combination of disk viscosity $\alpha$ and the Eddington fraction $f_{\rm Edd}$~\cite{speriProbingAccretionPhysics2023}. 
However,
unless other observations (in addition to the GW ones) are available to break the degeneracy between $\alpha$ and $f_{\rm Edd}$, extracting the local properties of the disk (e.g. its surface density) will be challenging for circular orbits. This degeneracy can be broken when eccentricity is included, cf. e.g.~\cite{Duque:2024mfw}.
Alternatively, one may measure the Eddington ratio
from electromagnetic observations of the host AGN, provided that the latter can be identified in the sky errorbox returned by GW observations \cite{speriProbingAccretionPhysics2023}
Besides being degenerate, the relation between $A$ and $\alpha,\,f_{\rm Edd}$ 
depends sensitively on the disk model, which is subject to systematic uncertainties. This relation is especially hard to model accurately in the Type II regime, as discussed in Sec.~\ref{sec:plan_mig}. This can cause additional biases in the astrophysical inference.

\begin{figure*}
    \centering
    \includegraphics[scale=0.5]{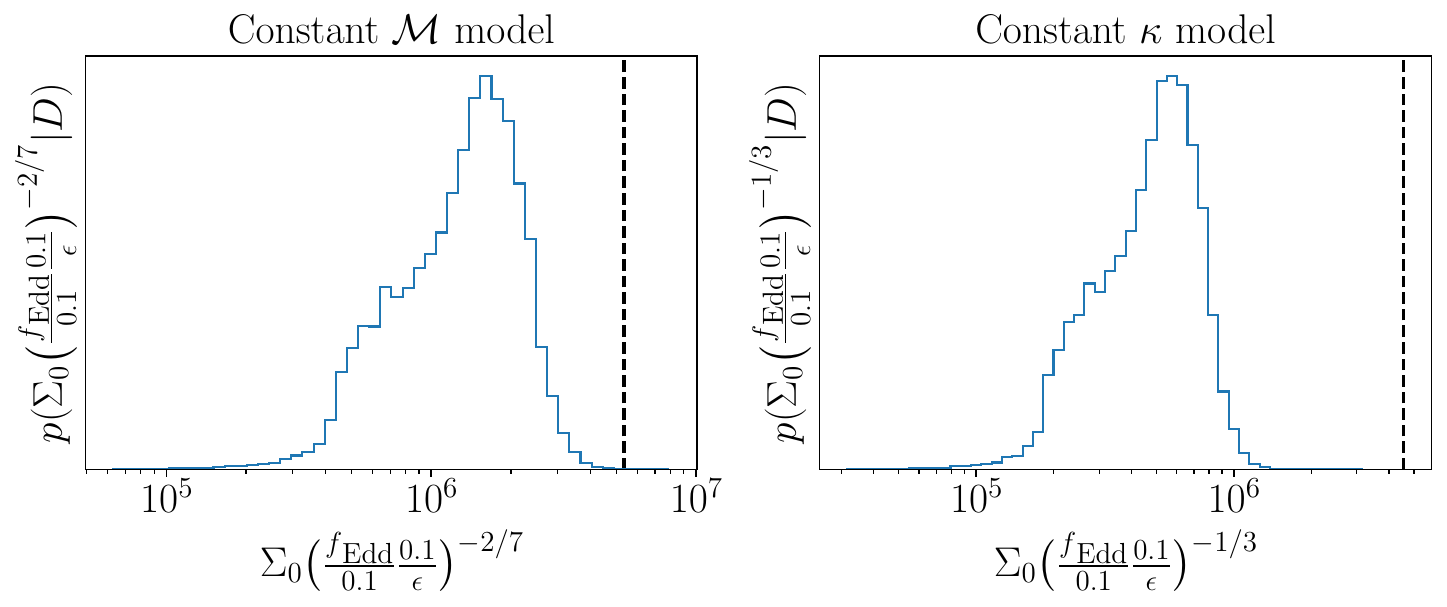}
    \caption{
    Disk properties inferred from gravitational wave observations for case 1, using two different disk models.
    In the {\it left panel}, we report the posterior obtained using a constant Mach number $\mathcal{M}$ disk model, matching the one used in the simulations.
    In black, we show the true value, obtained from the density $\Sigma_0$ and radiative efficiency $\epsilon$ used in the simulations and $f\textsubscript{Edd}$ obtained from Eq.~\eqref{eq:sigma_Mconst} ($f\textsubscript{Edd} = 287$).
    Even when using the correct disk model, we are unable to accurately infer the properties of the disk due to the stochastic fluctuations in the torque. 
    In the {\it right panel}, we repeat the analysis with a constant opacity $\kappa$ disk model.
    While this choice is compatible with the torque slope inferred from the signal, it does not match the disk model used in the simulations, thus introducing further a bias.}
    \label{fig:alpha_fedd_posterior}
\end{figure*}

We model the amplitude of the torque assuming a thin isothermal disk with the $\alpha$ viscosity prescription and assuming negligible radiative cooling. 
Solving the equilibrium equations for a Newtonian disk~\cite{frankAccretionPowerAstrophysics2002},
we find that  an opacity prescription $\kappa \propto \rho T_c ^{-1}$ leads to a steady state disk with constant Mach number:
\begin{align}
 \Sigma &= 1.5\times 10^6 M_6 ^{4/15} \alpha_{0.01} ^{-11/15} f_{0.1}^{3/5} r_{10}^{-1/2} ~\si{\kilo\gram\per \metre^2}
 \label{eq:sigma_Mconst}\\
 H &=4.2\times 10^{8} M_6 ^{13/15} f_{0.1}^{1/5}r_{10} ~\si{\metre}\,,
 \label{eq:A_Mconst}
\end{align}
where $M_6 = M_1/10^6M_\odot$, $\alpha_{0.01}=\alpha/0.01$, $f_{0.1}=(f\textsubscript{Edd}/0.1)$, $r_{10}=r/10~M_1$.
For constant opacity $\kappa = 0.22~\si{\centi\metre^2/\gram}$ we find instead 
\begin{align}
 \Sigma &= 1.1\times 10^5 M_6 ^{1/5} \alpha_{0.01} ^{-4/5} f_{0.1}^{38/5} r_{10}^{-3/5} ~\si{\kilo\gram\per \metre^2}
 \label{eq:sigma_kconst}\\
 H &=9.6\times 10^{7} M_6 ^{1/10} f_{0.1}^{1/5}r_{10} ^{21/20}~\si{\metre}
 \label{eq:A_kconst}
\end{align}
As already mentioned,
this latter model is slightly different from the one used in the simulations from which we extracted the torque~\cite{derdzinskiEvolutionGasDisc2021,derdzinskiProbingGasDisc2019}, as the resulting Mach number is not constant across the disk, but rather evolves as $\mathcal{M}\propto r^{-1/20}$, and the expected value of the power-law slope is $n=22/5\simeq 4.4$ rather than $n=4$.
However, these values are still well within the $2\sigma$ posteriors, and thus very well compatible with the results of the analysis.

Based on the GW parameter estimation alone, assuming the analytic torque amplitude given by Eq.~\eqref{eq:Tanaka_torque} and the disk models described above, one can infer the posterior of $\Sigma_0(f\textsubscript{Edd}/ \epsilon)^{-2/7}$ for the constant Mach number disk model, and $\Sigma_0(f\textsubscript{Edd}/\epsilon)^{-1/3}$ for the constant opacity one.
We compare the  posteriors inferred for case 1 with the values used in the simulation ($\alpha=0.01$ with $\epsilon = 0.1$).
The true value of  $f\textsubscript{Edd}$ instead is obtained from Eq.~\eqref{eq:sigma_Mconst}.

As one can see in the left panel of  Fig.~\ref{fig:alpha_fedd_posterior}, for the constant Mach number disk model the recovered posterior already presents a bias of over two sigmas.
This happens because in our analysis we are unable to confidently distinguish events undergoing stochastic type II migration from ones undergoing simple Type I migration. 
For case 2, for example, we have seen that the posteriors produced by injecting a model undergoing Type I migration and one undergoing gap-opening migration are not significantly different.
Since the posteriors on $A$ depend, in the stochastic case, on the averaged net torque, for which an exact relation linking it to the disk properties is unknown, the inference of the environment properties can result in biases even when knowing the correct disk model.

In the right panel of Fig.~\ref{fig:alpha_fedd_posterior}, we also report the disk properties inferred using the   constant $\kappa$ disk model.
We see here that the bias, already present due to the stochastic features in the torque, is  amplified by the choice of the wrong disk model.

\section{Conclusion}
 In this study, we investigated the robustness of simple analytic models of gas torques to perform parameter estimation for GW signals from EMRIs/IMRIs in accretion disks, as detectable by LISA. Using hydrodynamic simulations of accretion disk torques as injections, we investigated the effect of stochastic torque variability on the inference of both binary and environmental parameters using analytic templates. 
 Our parameter estimation studies demonstrate that, while simple torque models without stochastic variability allow for accurate recovery of binary parameters, they can bias the inferred environmental effects and accretion disk properties.  For example, the EMRI system of case 2 shows that masses, spins, and orbital separation are recovered accurately and precisely even with an analytic torque model as shown in Fig. 4.

For moderate stochastic time variability, the posteriors for the torque amplitude are approximately centered on the averaged value of the simulated torque. However, for significant stochastic variability, the estimated  amplitude may effectively reduce to zero, thus obscuring the presence of torques. This result underscores the subtleties that one has to face when inferring accretion disk properties from LISA observations using simplified torque models. While these models are sufficient for binary parameter extraction, a more sophisticated approach may be needed to reliably constrain the properties of the astrophysical environment. 

Indeed, even in the presence of unbiased estimation of the torque amplitude, we have shown that converting the latter to estimates of the disk's physical properties is subject to additional uncertainties, due to the possible mismatch between the disk model used in the analysis and the one describing the real astrophysical system (or, in our case, the simulations). 
Moreover, even when the bias in the recovered torque amplitudes is absent, the stochastic components of the torques can enhance the residual power left in the data.
Although these residuals are small (with SNR $\lesssim 1$) and should therefore not pose a problem for the LISA global fit,
their combined effect if thousands of EMRI sources are present in the data~\cite{babakScienceSpacebasedInterferometer2017}
may warrant further investigation.

In this work, we have pushed the mass ratio of the binary well within the IMRI range $q\sim 10^{-3}$, where nonlinear interactions between the disk and the secondary enhance the stochastic component of the gas torques. We have also considered relatively large surface densities near the disk's edge, $\Sigma\sim 10^7-10^8 \mbox{kg}/\mbox{m}^2$. On the other hand, the simulations that we consider adopt Mach numbers lower than the realistic value of $O(100)$. Boosting the Mach number may enhance stochastic torques compared to current simulations. Moreover, although for smaller  mass ratios (EMRIs) one may expect nonlinearities to play a less prominent role, turbulence might become the dominant source of stochastic fluctuations for those systems. More simulations will therefore be needed to assess torque fluctuations across the EMRI/IMRI and disk parameter space. 
 In some regimes of parameter space, periodicity is observed in gas torques, which leads to additional opportunities to measure disk signatures in the GW signal \cite{Zwick2024}.

Possible future improvements of this work include  three key areas: simulations, modeling and data analysis.
\textbf{Simulations}: 
Current Newtonian hydrodynamic simulations of accretion disks adopt idealized conditions (e.g., fixed Mach numbers, simplified viscosity). Future work should prioritize physically realistic models that incorporate high Mach numbers (\(\mathcal{M} \sim 100\)), radiative cooling, and magnetohydrodynamic turbulence.
Currently, relativistic magnetohydrodynamic disk simulations exist for disks MBH binaries \cite{Ennoggi2025,Avara2024,Combi2022,Gutierrez2022}, but they are limited in their evolution timescales and the parameter space towards extreme mass ratios, due to the computational resources required.
Additionally, 
general-relativistic disk models and/or different density profiles may impact the torques, see e.g.~\cite{shapiro_accretion_2013}.
There is also an increasing number of simpler hydrodynamic models that explore longer term evolution of gas around E/IMRIs \cite{Peng:2024wqf, PengChen2023, Clyburn2024, Sanchez2020}. Even incremental enhancements in physical prescriptions may better reproduce the stochastic torque variability observed in astrophysical systems, enabling more accurate prescriptions for torque modeling, and bridging the gap between numerical results and waveform templates.

\textbf{Waveform Modeling}:
While analytic torque models suffice for the systems studied here, their applicability across the full parameter space -- e.g. to eccentric or inclined orbits -- remains untested. Future waveform models could incorporate stochastic torque components through mathematical frameworks like stochastic differential equations, which could describe torque-driven orbital evolution. Hybrid approaches, blending analytic foundations with empirical corrections calibrated to simulations, may balance physical fidelity with computational tractability.

\textbf{Data Analysis}: 
The LISA data stream will contain overlapping signals from multiple sources, raising questions about how environmental torque signatures (or their residuals) might bias source recovery. While our analysis found negligible residual SNRs for isolated EMRI systems, future studies must assess whether the presence of other sources can impact the investigation of massive black hole environments (and vice versa). Developing noise-robust pipelines to disentangle environmental effects from confusion noise will be critical for reliable inference.

Progress in simulations, waveform modeling, and data analysis could potentially enable LISA observations to disentangle accretion disk physics from gravitational-wave signals, transforming accretion torques into a measurable probe of black hole environments.

\section*{Acknowledgements}
 E.B. and L.C. acknowledge support from the European Union’s H2020 ERC Consolidator Grant ``GRavity from Astrophysical to Microscopic Scales'' (Grant No. GRAMS-815673), the European Union’s Horizon  
ERC Synergy Grant ``Making Sense of the Unexpected in the Gravitational-Wave Sky'' (Grant No. GWSky-101167314), the PRIN 2022 grant ``GUVIRP - Gravity tests in the UltraViolet and InfraRed with Pulsar timing'', and the EU Horizon 2020 Research and Innovation Programme under the Marie Sklodowska-Curie Grant Agreement No. 101007855. L.S. acknowledges support from the UKRI guarantee funding (project no. EP/Y023706/1) and by the University of Nottingham Anne McLaren Fellowship.
This work has been supported by the
Agenzia Spaziale Italiana (ASI), Project n. 2024-36-HH.0, ``Attività per
la fase B2/C della missione LISA''.
Computational work have been made possible through SISSA-CINECA agreements providing access to resources on LEONARDO at CINECA.MB.

\bibliographystyle{apsrev}
\bibliography{Biblio} 

\begin{thebibliography}{109}
\expandafter\ifx\csname natexlab\endcsname\relax\def\natexlab#1{#1}\fi
\expandafter\ifx\csname bibnamefont\endcsname\relax
  \def\bibnamefont#1{#1}\fi
\expandafter\ifx\csname bibfnamefont\endcsname\relax
  \def\bibfnamefont#1{#1}\fi
\expandafter\ifx\csname citenamefont\endcsname\relax
  \def\citenamefont#1{#1}\fi
\expandafter\ifx\csname url\endcsname\relax
  \def\url#1{\texttt{#1}}\fi
\expandafter\ifx\csname urlprefix\endcsname\relax\def\urlprefix{URL }\fi
\providecommand{\bibinfo}[2]{#2}
\providecommand{\eprint}[2][]{\url{#2}}

\bibitem[{\citenamefont{Abbott et~al.}(2023{\natexlab{a}})\citenamefont{Abbott,
  Abbott, Acernese, Ackley, Adams, Adhikari, Adhikari, Adya, Affeldt, Agarwal
  et~al.}}]{PhysRevX.13.041039}
\bibinfo{author}{\bibfnamefont{R.}~\bibnamefont{Abbott}},
  \bibinfo{author}{\bibfnamefont{T.~D.} \bibnamefont{Abbott}},
  \bibinfo{author}{\bibfnamefont{F.}~\bibnamefont{Acernese}},
  \bibinfo{author}{\bibfnamefont{K.}~\bibnamefont{Ackley}},
  \bibinfo{author}{\bibfnamefont{C.}~\bibnamefont{Adams}},
  \bibinfo{author}{\bibfnamefont{N.}~\bibnamefont{Adhikari}},
  \bibinfo{author}{\bibfnamefont{R.~X.} \bibnamefont{Adhikari}},
  \bibinfo{author}{\bibfnamefont{V.~B.} \bibnamefont{Adya}},
  \bibinfo{author}{\bibfnamefont{C.}~\bibnamefont{Affeldt}},
  \bibinfo{author}{\bibfnamefont{D.}~\bibnamefont{Agarwal}},
  \bibnamefont{et~al.} (\bibinfo{collaboration}{LIGO Scientific Collaboration,
  Virgo Collaboration, and KAGRA Collaboration}), \bibinfo{journal}{Phys. Rev.
  X} \textbf{\bibinfo{volume}{13}}, \bibinfo{pages}{041039}
  (\bibinfo{year}{2023}{\natexlab{a}}),
  \urlprefix\url{https://link.aps.org/doi/10.1103/PhysRevX.13.041039}.

\bibitem[{\citenamefont{Abbott et~al.}(2024)\citenamefont{Abbott, Abbott,
  Acernese, Ackley, Adams, Adhikari, Adhikari, Adya, Affeldt, Agarwal
  et~al.}}]{gwtc2:1}
\bibinfo{author}{\bibfnamefont{R.}~\bibnamefont{Abbott}},
  \bibinfo{author}{\bibfnamefont{T.~D.} \bibnamefont{Abbott}},
  \bibinfo{author}{\bibfnamefont{F.}~\bibnamefont{Acernese}},
  \bibinfo{author}{\bibfnamefont{K.}~\bibnamefont{Ackley}},
  \bibinfo{author}{\bibfnamefont{C.}~\bibnamefont{Adams}},
  \bibinfo{author}{\bibfnamefont{N.}~\bibnamefont{Adhikari}},
  \bibinfo{author}{\bibfnamefont{R.~X.} \bibnamefont{Adhikari}},
  \bibinfo{author}{\bibfnamefont{V.~B.} \bibnamefont{Adya}},
  \bibinfo{author}{\bibfnamefont{C.}~\bibnamefont{Affeldt}},
  \bibinfo{author}{\bibfnamefont{D.}~\bibnamefont{Agarwal}},
  \bibnamefont{et~al.} (\bibinfo{collaboration}{The LIGO Scientific
  Collaboration and the Virgo Collaboration}), \bibinfo{journal}{Phys. Rev. D}
  \textbf{\bibinfo{volume}{109}}, \bibinfo{pages}{022001}
  (\bibinfo{year}{2024}),
  \urlprefix\url{https://link.aps.org/doi/10.1103/PhysRevD.109.022001}.

\bibitem[{\citenamefont{Abbott et~al.}(2021{\natexlab{a}})\citenamefont{Abbott,
  Abbott, Abraham, Acernese, Ackley, Adams, Adams, Adhikari, Adya, Affeldt
  et~al.}}]{gwtc2}
\bibinfo{author}{\bibfnamefont{R.}~\bibnamefont{Abbott}},
  \bibinfo{author}{\bibfnamefont{T.~D.} \bibnamefont{Abbott}},
  \bibinfo{author}{\bibfnamefont{S.}~\bibnamefont{Abraham}},
  \bibinfo{author}{\bibfnamefont{F.}~\bibnamefont{Acernese}},
  \bibinfo{author}{\bibfnamefont{K.}~\bibnamefont{Ackley}},
  \bibinfo{author}{\bibfnamefont{A.}~\bibnamefont{Adams}},
  \bibinfo{author}{\bibfnamefont{C.}~\bibnamefont{Adams}},
  \bibinfo{author}{\bibfnamefont{R.~X.} \bibnamefont{Adhikari}},
  \bibinfo{author}{\bibfnamefont{V.~B.} \bibnamefont{Adya}},
  \bibinfo{author}{\bibfnamefont{C.}~\bibnamefont{Affeldt}},
  \bibnamefont{et~al.} (\bibinfo{collaboration}{LIGO Scientific Collaboration
  and Virgo Collaboration}), \bibinfo{journal}{Phys. Rev. X}
  \textbf{\bibinfo{volume}{11}}, \bibinfo{pages}{021053}
  (\bibinfo{year}{2021}{\natexlab{a}}),
  \urlprefix\url{https://link.aps.org/doi/10.1103/PhysRevX.11.021053}.

\bibitem[{\citenamefont{Abbott et~al.}(2019{\natexlab{a}})\citenamefont{Abbott,
  Abbott, Abbott, Abraham, Acernese, Ackley, Adams, Adhikari, Adya, Affeldt
  et~al.}}]{gwtc1}
\bibinfo{author}{\bibfnamefont{B.~P.} \bibnamefont{Abbott}},
  \bibinfo{author}{\bibfnamefont{R.}~\bibnamefont{Abbott}},
  \bibinfo{author}{\bibfnamefont{T.~D.} \bibnamefont{Abbott}},
  \bibinfo{author}{\bibfnamefont{S.}~\bibnamefont{Abraham}},
  \bibinfo{author}{\bibfnamefont{F.}~\bibnamefont{Acernese}},
  \bibinfo{author}{\bibfnamefont{K.}~\bibnamefont{Ackley}},
  \bibinfo{author}{\bibfnamefont{C.}~\bibnamefont{Adams}},
  \bibinfo{author}{\bibfnamefont{R.~X.} \bibnamefont{Adhikari}},
  \bibinfo{author}{\bibfnamefont{V.~B.} \bibnamefont{Adya}},
  \bibinfo{author}{\bibfnamefont{C.}~\bibnamefont{Affeldt}},
  \bibnamefont{et~al.} (\bibinfo{collaboration}{LIGO Scientific Collaboration
  and Virgo Collaboration}), \bibinfo{journal}{Phys. Rev. X}
  \textbf{\bibinfo{volume}{9}}, \bibinfo{pages}{031040}
  (\bibinfo{year}{2019}{\natexlab{a}}),
  \urlprefix\url{https://link.aps.org/doi/10.1103/PhysRevX.9.031040}.

\bibitem[{\citenamefont{Abbott et~al.}(2016{\natexlab{a}})\citenamefont{Abbott,
  Abbott, Abbott, Abernathy, Acernese, Ackley, Adams, Adams, Addesso, Adhikari
  et~al.}}]{PhysRevLett.116.061102}
\bibinfo{author}{\bibfnamefont{B.~P.} \bibnamefont{Abbott}},
  \bibinfo{author}{\bibfnamefont{R.}~\bibnamefont{Abbott}},
  \bibinfo{author}{\bibfnamefont{T.~D.} \bibnamefont{Abbott}},
  \bibinfo{author}{\bibfnamefont{M.~R.} \bibnamefont{Abernathy}},
  \bibinfo{author}{\bibfnamefont{F.}~\bibnamefont{Acernese}},
  \bibinfo{author}{\bibfnamefont{K.}~\bibnamefont{Ackley}},
  \bibinfo{author}{\bibfnamefont{C.}~\bibnamefont{Adams}},
  \bibinfo{author}{\bibfnamefont{T.}~\bibnamefont{Adams}},
  \bibinfo{author}{\bibfnamefont{P.}~\bibnamefont{Addesso}},
  \bibinfo{author}{\bibfnamefont{R.~X.} \bibnamefont{Adhikari}},
  \bibnamefont{et~al.} (\bibinfo{collaboration}{LIGO Scientific Collaboration
  and Virgo Collaboration}), \bibinfo{journal}{Phys. Rev. Lett.}
  \textbf{\bibinfo{volume}{116}}, \bibinfo{pages}{061102}
  (\bibinfo{year}{2016}{\natexlab{a}}),
  \urlprefix\url{https://link.aps.org/doi/10.1103/PhysRevLett.116.061102}.

\bibitem[{\citenamefont{Abbott et~al.}(2017)\citenamefont{Abbott, Abbott,
  Abbott, Acernese, Ackley, Adams, Adams, Addesso, Adhikari, Adya
  et~al.}}]{GW170817}
\bibinfo{author}{\bibfnamefont{B.~P.} \bibnamefont{Abbott}},
  \bibinfo{author}{\bibfnamefont{R.}~\bibnamefont{Abbott}},
  \bibinfo{author}{\bibfnamefont{T.~D.} \bibnamefont{Abbott}},
  \bibinfo{author}{\bibfnamefont{F.}~\bibnamefont{Acernese}},
  \bibinfo{author}{\bibfnamefont{K.}~\bibnamefont{Ackley}},
  \bibinfo{author}{\bibfnamefont{C.}~\bibnamefont{Adams}},
  \bibinfo{author}{\bibfnamefont{T.}~\bibnamefont{Adams}},
  \bibinfo{author}{\bibfnamefont{P.}~\bibnamefont{Addesso}},
  \bibinfo{author}{\bibfnamefont{R.~X.} \bibnamefont{Adhikari}},
  \bibinfo{author}{\bibfnamefont{V.~B.} \bibnamefont{Adya}},
  \bibnamefont{et~al.} (\bibinfo{collaboration}{LIGO Scientific Collaboration
  and Virgo Collaboration}), \bibinfo{journal}{Phys. Rev. Lett.}
  \textbf{\bibinfo{volume}{119}}, \bibinfo{pages}{161101}
  (\bibinfo{year}{2017}),
  \urlprefix\url{https://link.aps.org/doi/10.1103/PhysRevLett.119.161101}.

\bibitem[{\citenamefont{Abbott et~al.}(2023{\natexlab{b}})}]{KAGRA:2021duu}
\bibinfo{author}{\bibfnamefont{R.}~\bibnamefont{Abbott}} \bibnamefont{et~al.}
  (\bibinfo{collaboration}{KAGRA, VIRGO, LIGO Scientific}),
  \bibinfo{journal}{Phys. Rev. X} \textbf{\bibinfo{volume}{13}},
  \bibinfo{pages}{011048} (\bibinfo{year}{2023}{\natexlab{b}}),
  \eprint{2111.03634}.

\bibitem[{\citenamefont{Abbott et~al.}(2019{\natexlab{b}})\citenamefont{Abbott,
  Abbott, Abbott, Abraham, Acernese, Ackley, Adams, Adhikari, Adya, Affeldt
  et~al.}}]{Abbott_2019}
\bibinfo{author}{\bibfnamefont{B.~P.} \bibnamefont{Abbott}},
  \bibinfo{author}{\bibfnamefont{R.}~\bibnamefont{Abbott}},
  \bibinfo{author}{\bibfnamefont{T.~D.} \bibnamefont{Abbott}},
  \bibinfo{author}{\bibfnamefont{S.}~\bibnamefont{Abraham}},
  \bibinfo{author}{\bibfnamefont{F.}~\bibnamefont{Acernese}},
  \bibinfo{author}{\bibfnamefont{K.}~\bibnamefont{Ackley}},
  \bibinfo{author}{\bibfnamefont{C.}~\bibnamefont{Adams}},
  \bibinfo{author}{\bibfnamefont{R.~X.} \bibnamefont{Adhikari}},
  \bibinfo{author}{\bibfnamefont{V.~B.} \bibnamefont{Adya}},
  \bibinfo{author}{\bibfnamefont{C.}~\bibnamefont{Affeldt}},
  \bibnamefont{et~al.}, \bibinfo{journal}{The Astrophysical Journal Letters}
  \textbf{\bibinfo{volume}{882}}, \bibinfo{pages}{L24}
  (\bibinfo{year}{2019}{\natexlab{b}}),
  \urlprefix\url{https://dx.doi.org/10.3847/2041-8213/ab3800}.

\bibitem[{\citenamefont{Abbott et~al.}(2021{\natexlab{b}})\citenamefont{Abbott,
  Abbott, Abraham, Acernese, Ackley, Adams, Adams, Adhikari, Adya, Affeldt
  et~al.}}]{Abbott_2021}
\bibinfo{author}{\bibfnamefont{R.}~\bibnamefont{Abbott}},
  \bibinfo{author}{\bibfnamefont{T.~D.} \bibnamefont{Abbott}},
  \bibinfo{author}{\bibfnamefont{S.}~\bibnamefont{Abraham}},
  \bibinfo{author}{\bibfnamefont{F.}~\bibnamefont{Acernese}},
  \bibinfo{author}{\bibfnamefont{K.}~\bibnamefont{Ackley}},
  \bibinfo{author}{\bibfnamefont{A.}~\bibnamefont{Adams}},
  \bibinfo{author}{\bibfnamefont{C.}~\bibnamefont{Adams}},
  \bibinfo{author}{\bibfnamefont{R.~X.} \bibnamefont{Adhikari}},
  \bibinfo{author}{\bibfnamefont{V.~B.} \bibnamefont{Adya}},
  \bibinfo{author}{\bibfnamefont{C.}~\bibnamefont{Affeldt}},
  \bibnamefont{et~al.}, \bibinfo{journal}{The Astrophysical Journal Letters}
  \textbf{\bibinfo{volume}{913}}, \bibinfo{pages}{L7}
  (\bibinfo{year}{2021}{\natexlab{b}}),
  \urlprefix\url{https://dx.doi.org/10.3847/2041-8213/abe949}.

\bibitem[{\citenamefont{Abbott et~al.}(2016{\natexlab{b}})\citenamefont{Abbott,
  Abbott, Abbott, Abernathy, Acernese, Ackley, Adams, Adams, Addesso, Adhikari
  et~al.}}]{Fchannel:GW150914}
\bibinfo{author}{\bibfnamefont{B.~P.} \bibnamefont{Abbott}},
  \bibinfo{author}{\bibfnamefont{R.}~\bibnamefont{Abbott}},
  \bibinfo{author}{\bibfnamefont{T.~D.} \bibnamefont{Abbott}},
  \bibinfo{author}{\bibfnamefont{M.~R.} \bibnamefont{Abernathy}},
  \bibinfo{author}{\bibfnamefont{F.}~\bibnamefont{Acernese}},
  \bibinfo{author}{\bibfnamefont{K.}~\bibnamefont{Ackley}},
  \bibinfo{author}{\bibfnamefont{C.}~\bibnamefont{Adams}},
  \bibinfo{author}{\bibfnamefont{T.}~\bibnamefont{Adams}},
  \bibinfo{author}{\bibfnamefont{P.}~\bibnamefont{Addesso}},
  \bibinfo{author}{\bibfnamefont{R.~X.} \bibnamefont{Adhikari}},
  \bibnamefont{et~al.}, \bibinfo{journal}{The Astrophysical Journal Letters}
  \textbf{\bibinfo{volume}{818}}, \bibinfo{pages}{L22}
  (\bibinfo{year}{2016}{\natexlab{b}}),
  \urlprefix\url{https://dx.doi.org/10.3847/2041-8205/818/2/L22}.

\bibitem[{\citenamefont{Abbott et~al.}(2020{\natexlab{a}})\citenamefont{Abbott,
  Abbott, Abraham, Acernese, Ackley, Adams, Adhikari, Adya, Affeldt, Agathos
  et~al.}}]{Fchannel:GW190521}
\bibinfo{author}{\bibfnamefont{R.}~\bibnamefont{Abbott}},
  \bibinfo{author}{\bibfnamefont{T.~D.} \bibnamefont{Abbott}},
  \bibinfo{author}{\bibfnamefont{S.}~\bibnamefont{Abraham}},
  \bibinfo{author}{\bibfnamefont{F.}~\bibnamefont{Acernese}},
  \bibinfo{author}{\bibfnamefont{K.}~\bibnamefont{Ackley}},
  \bibinfo{author}{\bibfnamefont{C.}~\bibnamefont{Adams}},
  \bibinfo{author}{\bibfnamefont{R.~X.} \bibnamefont{Adhikari}},
  \bibinfo{author}{\bibfnamefont{V.~B.} \bibnamefont{Adya}},
  \bibinfo{author}{\bibfnamefont{C.}~\bibnamefont{Affeldt}},
  \bibinfo{author}{\bibfnamefont{M.}~\bibnamefont{Agathos}},
  \bibnamefont{et~al.}, \bibinfo{journal}{The Astrophysical Journal Letters}
  \textbf{\bibinfo{volume}{900}}, \bibinfo{pages}{L13}
  (\bibinfo{year}{2020}{\natexlab{a}}),
  \urlprefix\url{https://dx.doi.org/10.3847/2041-8213/aba493}.

\bibitem[{\citenamefont{Collaboration et~al.}(2021)\citenamefont{Collaboration,
  the Virgo~Collaboration, the KAGRA~Collaboration, Abbott, Abe, Acernese,
  Ackley, Adhikari, Adhikari, Adkins
  et~al.}}]{theligoscientificcollaboration2021testsgeneralrelativitygwtc3}
\bibinfo{author}{\bibfnamefont{T.~L.~S.} \bibnamefont{Collaboration}},
  \bibinfo{author}{\bibnamefont{the Virgo~Collaboration}},
  \bibinfo{author}{\bibnamefont{the KAGRA~Collaboration}},
  \bibinfo{author}{\bibfnamefont{R.}~\bibnamefont{Abbott}},
  \bibinfo{author}{\bibfnamefont{H.}~\bibnamefont{Abe}},
  \bibinfo{author}{\bibfnamefont{F.}~\bibnamefont{Acernese}},
  \bibinfo{author}{\bibfnamefont{K.}~\bibnamefont{Ackley}},
  \bibinfo{author}{\bibfnamefont{N.}~\bibnamefont{Adhikari}},
  \bibinfo{author}{\bibfnamefont{R.~X.} \bibnamefont{Adhikari}},
  \bibinfo{author}{\bibfnamefont{V.~K.} \bibnamefont{Adkins}},
  \bibnamefont{et~al.}, \emph{\bibinfo{title}{Tests of general relativity with
  gwtc-3}} (\bibinfo{year}{2021}), \eprint{2112.06861},
  \urlprefix\url{https://arxiv.org/abs/2112.06861}.

\bibitem[{\citenamefont{Abbott et~al.}(2016{\natexlab{c}})\citenamefont{Abbott,
  Abbott, Abbott, Abernathy, Acernese, Ackley, Adams, Adams, Addesso, Adhikari
  et~al.}}]{TestGr:GW150914}
\bibinfo{author}{\bibfnamefont{B.~P.} \bibnamefont{Abbott}},
  \bibinfo{author}{\bibfnamefont{R.}~\bibnamefont{Abbott}},
  \bibinfo{author}{\bibfnamefont{T.~D.} \bibnamefont{Abbott}},
  \bibinfo{author}{\bibfnamefont{M.~R.} \bibnamefont{Abernathy}},
  \bibinfo{author}{\bibfnamefont{F.}~\bibnamefont{Acernese}},
  \bibinfo{author}{\bibfnamefont{K.}~\bibnamefont{Ackley}},
  \bibinfo{author}{\bibfnamefont{C.}~\bibnamefont{Adams}},
  \bibinfo{author}{\bibfnamefont{T.}~\bibnamefont{Adams}},
  \bibinfo{author}{\bibfnamefont{P.}~\bibnamefont{Addesso}},
  \bibinfo{author}{\bibfnamefont{R.~X.} \bibnamefont{Adhikari}},
  \bibnamefont{et~al.} (\bibinfo{collaboration}{LIGO Scientific and Virgo
  Collaborations}), \bibinfo{journal}{Phys. Rev. Lett.}
  \textbf{\bibinfo{volume}{116}}, \bibinfo{pages}{221101}
  (\bibinfo{year}{2016}{\natexlab{c}}),
  \urlprefix\url{https://link.aps.org/doi/10.1103/PhysRevLett.116.221101}.

\bibitem[{\citenamefont{Abbott et~al.}(2019{\natexlab{c}})\citenamefont{Abbott,
  Abbott, Abbott, Acernese, Ackley, Adams, Adams, Addesso, Adhikari, Adya
  et~al.}}]{TestGR:GW170817}
\bibinfo{author}{\bibfnamefont{B.~P.} \bibnamefont{Abbott}},
  \bibinfo{author}{\bibfnamefont{R.}~\bibnamefont{Abbott}},
  \bibinfo{author}{\bibfnamefont{T.~D.} \bibnamefont{Abbott}},
  \bibinfo{author}{\bibfnamefont{F.}~\bibnamefont{Acernese}},
  \bibinfo{author}{\bibfnamefont{K.}~\bibnamefont{Ackley}},
  \bibinfo{author}{\bibfnamefont{C.}~\bibnamefont{Adams}},
  \bibinfo{author}{\bibfnamefont{T.}~\bibnamefont{Adams}},
  \bibinfo{author}{\bibfnamefont{P.}~\bibnamefont{Addesso}},
  \bibinfo{author}{\bibfnamefont{R.~X.} \bibnamefont{Adhikari}},
  \bibinfo{author}{\bibfnamefont{V.~B.} \bibnamefont{Adya}},
  \bibnamefont{et~al.} (\bibinfo{collaboration}{LIGO Scientific Collaboration
  and Virgo Collaboration}), \bibinfo{journal}{Phys. Rev. Lett.}
  \textbf{\bibinfo{volume}{123}}, \bibinfo{pages}{011102}
  (\bibinfo{year}{2019}{\natexlab{c}}),
  \urlprefix\url{https://link.aps.org/doi/10.1103/PhysRevLett.123.011102}.

\bibitem[{\citenamefont{Abbott et~al.}(2019{\natexlab{d}})\citenamefont{Abbott,
  Abbott, Abbott, Abraham, Acernese, Ackley, Adams, Adhikari, Adya, Affeldt
  et~al.}}]{TestGR:GWTC1}
\bibinfo{author}{\bibfnamefont{B.~P.} \bibnamefont{Abbott}},
  \bibinfo{author}{\bibfnamefont{R.}~\bibnamefont{Abbott}},
  \bibinfo{author}{\bibfnamefont{T.~D.} \bibnamefont{Abbott}},
  \bibinfo{author}{\bibfnamefont{S.}~\bibnamefont{Abraham}},
  \bibinfo{author}{\bibfnamefont{F.}~\bibnamefont{Acernese}},
  \bibinfo{author}{\bibfnamefont{K.}~\bibnamefont{Ackley}},
  \bibinfo{author}{\bibfnamefont{C.}~\bibnamefont{Adams}},
  \bibinfo{author}{\bibfnamefont{R.~X.} \bibnamefont{Adhikari}},
  \bibinfo{author}{\bibfnamefont{V.~B.} \bibnamefont{Adya}},
  \bibinfo{author}{\bibfnamefont{C.}~\bibnamefont{Affeldt}},
  \bibnamefont{et~al.} (\bibinfo{collaboration}{The LIGO Scientific
  Collaboration and the Virgo Collaboration}), \bibinfo{journal}{Phys. Rev. D}
  \textbf{\bibinfo{volume}{100}}, \bibinfo{pages}{104036}
  (\bibinfo{year}{2019}{\natexlab{d}}),
  \urlprefix\url{https://link.aps.org/doi/10.1103/PhysRevD.100.104036}.

\bibitem[{\citenamefont{Abbott et~al.}(2021{\natexlab{c}})\citenamefont{Abbott,
  Abbott, Abraham, Acernese, Ackley, Adams, Adams, Adhikari, Adya, Affeldt
  et~al.}}]{TestGR:GWTC2}
\bibinfo{author}{\bibfnamefont{R.}~\bibnamefont{Abbott}},
  \bibinfo{author}{\bibfnamefont{T.~D.} \bibnamefont{Abbott}},
  \bibinfo{author}{\bibfnamefont{S.}~\bibnamefont{Abraham}},
  \bibinfo{author}{\bibfnamefont{F.}~\bibnamefont{Acernese}},
  \bibinfo{author}{\bibfnamefont{K.}~\bibnamefont{Ackley}},
  \bibinfo{author}{\bibfnamefont{A.}~\bibnamefont{Adams}},
  \bibinfo{author}{\bibfnamefont{C.}~\bibnamefont{Adams}},
  \bibinfo{author}{\bibfnamefont{R.~X.} \bibnamefont{Adhikari}},
  \bibinfo{author}{\bibfnamefont{V.~B.} \bibnamefont{Adya}},
  \bibinfo{author}{\bibfnamefont{C.}~\bibnamefont{Affeldt}},
  \bibnamefont{et~al.} (\bibinfo{collaboration}{LIGO Scientific Collaboration
  and Virgo Collaboration}), \bibinfo{journal}{Phys. Rev. D}
  \textbf{\bibinfo{volume}{103}}, \bibinfo{pages}{122002}
  (\bibinfo{year}{2021}{\natexlab{c}}),
  \urlprefix\url{https://link.aps.org/doi/10.1103/PhysRevD.103.122002}.

\bibitem[{\citenamefont{Abbott et~al.}(2020{\natexlab{b}})\citenamefont{Abbott,
  Abbott, Abraham, Acernese, Ackley, Adams, Adhikari, Adya, Affeldt, Agathos
  et~al.}}]{PhysRevLett.125.101102}
\bibinfo{author}{\bibfnamefont{R.}~\bibnamefont{Abbott}},
  \bibinfo{author}{\bibfnamefont{T.~D.} \bibnamefont{Abbott}},
  \bibinfo{author}{\bibfnamefont{S.}~\bibnamefont{Abraham}},
  \bibinfo{author}{\bibfnamefont{F.}~\bibnamefont{Acernese}},
  \bibinfo{author}{\bibfnamefont{K.}~\bibnamefont{Ackley}},
  \bibinfo{author}{\bibfnamefont{C.}~\bibnamefont{Adams}},
  \bibinfo{author}{\bibfnamefont{R.~X.} \bibnamefont{Adhikari}},
  \bibinfo{author}{\bibfnamefont{V.~B.} \bibnamefont{Adya}},
  \bibinfo{author}{\bibfnamefont{C.}~\bibnamefont{Affeldt}},
  \bibinfo{author}{\bibfnamefont{M.}~\bibnamefont{Agathos}},
  \bibnamefont{et~al.} (\bibinfo{collaboration}{LIGO Scientific Collaboration
  and Virgo Collaboration}), \bibinfo{journal}{Phys. Rev. Lett.}
  \textbf{\bibinfo{volume}{125}}, \bibinfo{pages}{101102}
  (\bibinfo{year}{2020}{\natexlab{b}}),
  \urlprefix\url{https://link.aps.org/doi/10.1103/PhysRevLett.125.101102}.

\bibitem[{\citenamefont{Woosley}(2017)}]{woosleyPulsationalPairinstabilitySupernovae2017}
\bibinfo{author}{\bibfnamefont{S.~E.} \bibnamefont{Woosley}},
  \bibinfo{journal}{The Astrophysical Journal} \textbf{\bibinfo{volume}{836}},
  \bibinfo{pages}{244} (\bibinfo{year}{2017}), ISSN \bibinfo{issn}{0004-637X,
  1538-4357}.

\bibitem[{\citenamefont{Farmer et~al.}(2019)\citenamefont{Farmer, Renzo,
  de~Mink, Marchant, and Justham}}]{Farmer_2019}
\bibinfo{author}{\bibfnamefont{R.}~\bibnamefont{Farmer}},
  \bibinfo{author}{\bibfnamefont{M.}~\bibnamefont{Renzo}},
  \bibinfo{author}{\bibfnamefont{S.~E.} \bibnamefont{de~Mink}},
  \bibinfo{author}{\bibfnamefont{P.}~\bibnamefont{Marchant}}, \bibnamefont{and}
  \bibinfo{author}{\bibfnamefont{S.}~\bibnamefont{Justham}},
  \bibinfo{journal}{The Astrophysical Journal} \textbf{\bibinfo{volume}{887}},
  \bibinfo{pages}{53} (\bibinfo{year}{2019}),
  \urlprefix\url{https://dx.doi.org/10.3847/1538-4357/ab518b}.

\bibitem[{\citenamefont{Tagawa et~al.}(2020)\citenamefont{Tagawa, Haiman, and
  Kocsis}}]{Tagawa_2020}
\bibinfo{author}{\bibfnamefont{H.}~\bibnamefont{Tagawa}},
  \bibinfo{author}{\bibfnamefont{Z.}~\bibnamefont{Haiman}}, \bibnamefont{and}
  \bibinfo{author}{\bibfnamefont{B.}~\bibnamefont{Kocsis}},
  \bibinfo{journal}{The Astrophysical Journal} \textbf{\bibinfo{volume}{898}},
  \bibinfo{pages}{25} (\bibinfo{year}{2020}),
  \urlprefix\url{https://dx.doi.org/10.3847/1538-4357/ab9b8c}.

\bibitem[{\citenamefont{Toubiana
  et~al.}(2021{\natexlab{a}})}]{Toubiana:2020drf}
\bibinfo{author}{\bibfnamefont{A.}~\bibnamefont{Toubiana}}
  \bibnamefont{et~al.}, \bibinfo{journal}{Phys. Rev. Lett.}
  \textbf{\bibinfo{volume}{126}}, \bibinfo{pages}{101105}
  (\bibinfo{year}{2021}{\natexlab{a}}), \eprint{2010.06056}.

\bibitem[{\citenamefont{Sberna et~al.}(2022{\natexlab{a}})}]{Sberna:2022qbn}
\bibinfo{author}{\bibfnamefont{L.}~\bibnamefont{Sberna}} \bibnamefont{et~al.},
  \bibinfo{journal}{Phys. Rev. D} \textbf{\bibinfo{volume}{106}},
  \bibinfo{pages}{064056} (\bibinfo{year}{2022}{\natexlab{a}}),
  \eprint{2205.08550}.

\bibitem[{\citenamefont{Caputo et~al.}(2020)\citenamefont{Caputo, Sberna,
  Toubiana, Babak, Barausse, Marsat, and
  Pani}}]{caputoGravitationalwaveDetectionParameter2020}
\bibinfo{author}{\bibfnamefont{A.}~\bibnamefont{Caputo}},
  \bibinfo{author}{\bibfnamefont{L.}~\bibnamefont{Sberna}},
  \bibinfo{author}{\bibfnamefont{A.}~\bibnamefont{Toubiana}},
  \bibinfo{author}{\bibfnamefont{S.}~\bibnamefont{Babak}},
  \bibinfo{author}{\bibfnamefont{E.}~\bibnamefont{Barausse}},
  \bibinfo{author}{\bibfnamefont{S.}~\bibnamefont{Marsat}}, \bibnamefont{and}
  \bibinfo{author}{\bibfnamefont{P.}~\bibnamefont{Pani}}, \bibinfo{journal}{The
  Astrophysical Journal} \textbf{\bibinfo{volume}{892}}, \bibinfo{pages}{90}
  (\bibinfo{year}{2020}), ISSN \bibinfo{issn}{1538-4357}, \eprint{2001.03620}.

\bibitem[{\citenamefont{Colpi et~al.}(2024)}]{Colpi:2024xhw}
\bibinfo{author}{\bibfnamefont{M.}~\bibnamefont{Colpi}} \bibnamefont{et~al.},
  \emph{\bibinfo{title}{{LISA Definition Study Report}}}
  (\bibinfo{year}{2024}), \eprint{2402.07571}.

\bibitem[{\citenamefont{Barausse and Rezzolla}(2008)}]{Barausse:2007dy}
\bibinfo{author}{\bibfnamefont{E.}~\bibnamefont{Barausse}} \bibnamefont{and}
  \bibinfo{author}{\bibfnamefont{L.}~\bibnamefont{Rezzolla}},
  \bibinfo{journal}{Phys. Rev. D} \textbf{\bibinfo{volume}{77}},
  \bibinfo{pages}{104027} (\bibinfo{year}{2008}), \eprint{0711.4558}.

\bibitem[{\citenamefont{Kocsis et~al.}(2011)\citenamefont{Kocsis, Yunes, and
  Loeb}}]{PhysRevD.84.024032}
\bibinfo{author}{\bibfnamefont{B.}~\bibnamefont{Kocsis}},
  \bibinfo{author}{\bibfnamefont{N.}~\bibnamefont{Yunes}}, \bibnamefont{and}
  \bibinfo{author}{\bibfnamefont{A.}~\bibnamefont{Loeb}},
  \bibinfo{journal}{Phys. Rev. D} \textbf{\bibinfo{volume}{84}},
  \bibinfo{pages}{024032} (\bibinfo{year}{2011}),
  \urlprefix\url{https://link.aps.org/doi/10.1103/PhysRevD.84.024032}.

\bibitem[{\citenamefont{Yunes et~al.}(2011)\citenamefont{Yunes, Kocsis, Loeb,
  and Haiman}}]{yunesImprintAccretionDiskInduced2011}
\bibinfo{author}{\bibfnamefont{N.}~\bibnamefont{Yunes}},
  \bibinfo{author}{\bibfnamefont{B.}~\bibnamefont{Kocsis}},
  \bibinfo{author}{\bibfnamefont{A.}~\bibnamefont{Loeb}}, \bibnamefont{and}
  \bibinfo{author}{\bibfnamefont{Z.}~\bibnamefont{Haiman}},
  \bibinfo{journal}{Physical Review Letters} \textbf{\bibinfo{volume}{107}},
  \bibinfo{pages}{171103} (\bibinfo{year}{2011}), ISSN
  \bibinfo{issn}{0031-9007, 1079-7114}, \eprint{1103.4609}.

\bibitem[{\citenamefont{Barausse et~al.}(2014)\citenamefont{Barausse, Cardoso,
  and Pani}}]{barausseCanEnvironmentalEffects2014}
\bibinfo{author}{\bibfnamefont{E.}~\bibnamefont{Barausse}},
  \bibinfo{author}{\bibfnamefont{V.}~\bibnamefont{Cardoso}}, \bibnamefont{and}
  \bibinfo{author}{\bibfnamefont{P.}~\bibnamefont{Pani}},
  \bibinfo{journal}{Physical Review D} \textbf{\bibinfo{volume}{89}},
  \bibinfo{pages}{104059} (\bibinfo{year}{2014}), ISSN
  \bibinfo{issn}{1550-7998, 1550-2368}, \eprint{1404.7149}.

\bibitem[{\citenamefont{Berry et~al.}(2019)\citenamefont{Berry, Hughes,
  Sopuerta, Chua, Heffernan, Holley-Bockelmann, Mihaylov, Miller, and
  Sesana}}]{berry2019uniquepotentialextrememassratio}
\bibinfo{author}{\bibfnamefont{C.~P.~L.} \bibnamefont{Berry}},
  \bibinfo{author}{\bibfnamefont{S.~A.} \bibnamefont{Hughes}},
  \bibinfo{author}{\bibfnamefont{C.~F.} \bibnamefont{Sopuerta}},
  \bibinfo{author}{\bibfnamefont{A.~J.~K.} \bibnamefont{Chua}},
  \bibinfo{author}{\bibfnamefont{A.}~\bibnamefont{Heffernan}},
  \bibinfo{author}{\bibfnamefont{K.}~\bibnamefont{Holley-Bockelmann}},
  \bibinfo{author}{\bibfnamefont{D.~P.} \bibnamefont{Mihaylov}},
  \bibinfo{author}{\bibfnamefont{M.~C.} \bibnamefont{Miller}},
  \bibnamefont{and} \bibinfo{author}{\bibfnamefont{A.}~\bibnamefont{Sesana}},
  \emph{\bibinfo{title}{The unique potential of extreme mass-ratio inspirals
  for gravitational-wave astronomy}} (\bibinfo{year}{2019}),
  \eprint{1903.03686}, \urlprefix\url{https://arxiv.org/abs/1903.03686}.

\bibitem[{\citenamefont{Toubiana
  et~al.}(2021{\natexlab{b}})\citenamefont{Toubiana, Sberna, Caputo, Cusin,
  Marsat, Jani, Babak, Barausse, Caprini, Pani
  et~al.}}]{toubianaDetectableEnvironmentalEffects2021}
\bibinfo{author}{\bibfnamefont{A.}~\bibnamefont{Toubiana}},
  \bibinfo{author}{\bibfnamefont{L.}~\bibnamefont{Sberna}},
  \bibinfo{author}{\bibfnamefont{A.}~\bibnamefont{Caputo}},
  \bibinfo{author}{\bibfnamefont{G.}~\bibnamefont{Cusin}},
  \bibinfo{author}{\bibfnamefont{S.}~\bibnamefont{Marsat}},
  \bibinfo{author}{\bibfnamefont{K.}~\bibnamefont{Jani}},
  \bibinfo{author}{\bibfnamefont{S.}~\bibnamefont{Babak}},
  \bibinfo{author}{\bibfnamefont{E.}~\bibnamefont{Barausse}},
  \bibinfo{author}{\bibfnamefont{C.}~\bibnamefont{Caprini}},
  \bibinfo{author}{\bibfnamefont{P.}~\bibnamefont{Pani}}, \bibnamefont{et~al.},
  \bibinfo{journal}{Physical Review Letters} \textbf{\bibinfo{volume}{126}},
  \bibinfo{pages}{101105} (\bibinfo{year}{2021}{\natexlab{b}}), ISSN
  \bibinfo{issn}{0031-9007, 1079-7114}, \eprint{2010.06056}.

\bibitem[{\citenamefont{Sberna et~al.}(2022{\natexlab{b}})\citenamefont{Sberna,
  Babak, Marsat, Caputo, Cusin, Toubiana, Barausse, Caprini, Canton, Sesana
  et~al.}}]{sbernaObservingGW190521likeBinary2022}
\bibinfo{author}{\bibfnamefont{L.}~\bibnamefont{Sberna}},
  \bibinfo{author}{\bibfnamefont{S.}~\bibnamefont{Babak}},
  \bibinfo{author}{\bibfnamefont{S.}~\bibnamefont{Marsat}},
  \bibinfo{author}{\bibfnamefont{A.}~\bibnamefont{Caputo}},
  \bibinfo{author}{\bibfnamefont{G.}~\bibnamefont{Cusin}},
  \bibinfo{author}{\bibfnamefont{A.}~\bibnamefont{Toubiana}},
  \bibinfo{author}{\bibfnamefont{E.}~\bibnamefont{Barausse}},
  \bibinfo{author}{\bibfnamefont{C.}~\bibnamefont{Caprini}},
  \bibinfo{author}{\bibfnamefont{T.~D.} \bibnamefont{Canton}},
  \bibinfo{author}{\bibfnamefont{A.}~\bibnamefont{Sesana}},
  \bibnamefont{et~al.}, \bibinfo{journal}{Physical Review D}
  \textbf{\bibinfo{volume}{106}}, \bibinfo{pages}{064056}
  (\bibinfo{year}{2022}{\natexlab{b}}), ISSN \bibinfo{issn}{2470-0010,
  2470-0029}, \eprint{2205.08550}.

\bibitem[{\citenamefont{Speri et~al.}(2023{\natexlab{a}})\citenamefont{Speri,
  Antonelli, Sberna, Babak, Barausse, Gair, and
  Katz}}]{speriProbingAccretionPhysics2023}
\bibinfo{author}{\bibfnamefont{L.}~\bibnamefont{Speri}},
  \bibinfo{author}{\bibfnamefont{A.}~\bibnamefont{Antonelli}},
  \bibinfo{author}{\bibfnamefont{L.}~\bibnamefont{Sberna}},
  \bibinfo{author}{\bibfnamefont{S.}~\bibnamefont{Babak}},
  \bibinfo{author}{\bibfnamefont{E.}~\bibnamefont{Barausse}},
  \bibinfo{author}{\bibfnamefont{J.~R.} \bibnamefont{Gair}}, \bibnamefont{and}
  \bibinfo{author}{\bibfnamefont{M.~L.} \bibnamefont{Katz}},
  \bibinfo{journal}{Phys. Rev. X} \textbf{\bibinfo{volume}{13}},
  \bibinfo{pages}{021035} (\bibinfo{year}{2023}{\natexlab{a}}),
  \eprint{2207.10086}.

\bibitem[{\citenamefont{{Garg} et~al.}(2024)\citenamefont{{Garg}, {Derdzinski},
  {Tiwari}, {Gair}, and {Mayer}}}]{Garg2024}
\bibinfo{author}{\bibfnamefont{M.}~\bibnamefont{{Garg}}},
  \bibinfo{author}{\bibfnamefont{A.}~\bibnamefont{{Derdzinski}}},
  \bibinfo{author}{\bibfnamefont{S.}~\bibnamefont{{Tiwari}}},
  \bibinfo{author}{\bibfnamefont{J.}~\bibnamefont{{Gair}}}, \bibnamefont{and}
  \bibinfo{author}{\bibfnamefont{L.}~\bibnamefont{{Mayer}}},
  \bibinfo{journal}{Monthly Notices of the Royal Astronomical Society}
  \textbf{\bibinfo{volume}{532}}, \bibinfo{pages}{4060} (\bibinfo{year}{2024}),
  \eprint{2402.14058}.

\bibitem[{\citenamefont{Babak et~al.}(2017)\citenamefont{Babak, Gair, Sesana,
  Barausse, Sopuerta, Berry, Berti, {Amaro-Seoane}, Petiteau, and
  Klein}}]{babakScienceSpacebasedInterferometer2017}
\bibinfo{author}{\bibfnamefont{S.}~\bibnamefont{Babak}},
  \bibinfo{author}{\bibfnamefont{J.}~\bibnamefont{Gair}},
  \bibinfo{author}{\bibfnamefont{A.}~\bibnamefont{Sesana}},
  \bibinfo{author}{\bibfnamefont{E.}~\bibnamefont{Barausse}},
  \bibinfo{author}{\bibfnamefont{C.~F.} \bibnamefont{Sopuerta}},
  \bibinfo{author}{\bibfnamefont{C.~P.~L.} \bibnamefont{Berry}},
  \bibinfo{author}{\bibfnamefont{E.}~\bibnamefont{Berti}},
  \bibinfo{author}{\bibfnamefont{P.}~\bibnamefont{{Amaro-Seoane}}},
  \bibinfo{author}{\bibfnamefont{A.}~\bibnamefont{Petiteau}}, \bibnamefont{and}
  \bibinfo{author}{\bibfnamefont{A.}~\bibnamefont{Klein}},
  \bibinfo{journal}{Physical Review D} \textbf{\bibinfo{volume}{95}},
  \bibinfo{pages}{103012} (\bibinfo{year}{2017}), ISSN
  \bibinfo{issn}{2470-0010, 2470-0029}, \eprint{1703.09722}.

\bibitem[{\citenamefont{Seoane et~al.}(2023)}]{LISA:2022yao}
\bibinfo{author}{\bibfnamefont{P.~A.} \bibnamefont{Seoane}}
  \bibnamefont{et~al.} (\bibinfo{collaboration}{LISA}),
  \bibinfo{journal}{Living Rev. Rel.} \textbf{\bibinfo{volume}{26}},
  \bibinfo{pages}{2} (\bibinfo{year}{2023}), \eprint{2203.06016}.

\bibitem[{\citenamefont{Greene et~al.}(2020)\citenamefont{Greene, Strader, and
  Ho}}]{Greene:2019vlv}
\bibinfo{author}{\bibfnamefont{J.~E.} \bibnamefont{Greene}},
  \bibinfo{author}{\bibfnamefont{J.}~\bibnamefont{Strader}}, \bibnamefont{and}
  \bibinfo{author}{\bibfnamefont{L.~C.} \bibnamefont{Ho}},
  \bibinfo{journal}{Ann. Rev. Astron. Astrophys.}
  \textbf{\bibinfo{volume}{58}}, \bibinfo{pages}{257} (\bibinfo{year}{2020}),
  \eprint{1911.09678}.

\bibitem[{\citenamefont{Kejriwal
  et~al.}(2024{\natexlab{a}})\citenamefont{Kejriwal, Witzany, Zajaček, Pasham,
  and Chua}}]{10.1093/mnras/stae1599}
\bibinfo{author}{\bibfnamefont{S.}~\bibnamefont{Kejriwal}},
  \bibinfo{author}{\bibfnamefont{V.}~\bibnamefont{Witzany}},
  \bibinfo{author}{\bibfnamefont{M.}~\bibnamefont{Zajaček}},
  \bibinfo{author}{\bibfnamefont{D.~R.} \bibnamefont{Pasham}},
  \bibnamefont{and} \bibinfo{author}{\bibfnamefont{A.~J.~K.}
  \bibnamefont{Chua}}, \bibinfo{journal}{Monthly Notices of the Royal
  Astronomical Society} \textbf{\bibinfo{volume}{532}}, \bibinfo{pages}{2143}
  (\bibinfo{year}{2024}{\natexlab{a}}), ISSN \bibinfo{issn}{0035-8711},
  \eprint{https://academic.oup.com/mnras/article-pdf/532/2/2143/58529727/stae1599.pdf},
  \urlprefix\url{https://doi.org/10.1093/mnras/stae1599}.

\bibitem[{\citenamefont{Pasham et~al.}(2024)\citenamefont{Pasham, Tombesi,
  Suková, Zajaček, Rakshit, Coughlin, Kosec, Karas, Masterson, Mummery
  et~al.}}]{doi:10.1126/sciadv.adj8898}
\bibinfo{author}{\bibfnamefont{D.~R.} \bibnamefont{Pasham}},
  \bibinfo{author}{\bibfnamefont{F.}~\bibnamefont{Tombesi}},
  \bibinfo{author}{\bibfnamefont{P.}~\bibnamefont{Suková}},
  \bibinfo{author}{\bibfnamefont{M.}~\bibnamefont{Zajaček}},
  \bibinfo{author}{\bibfnamefont{S.}~\bibnamefont{Rakshit}},
  \bibinfo{author}{\bibfnamefont{E.}~\bibnamefont{Coughlin}},
  \bibinfo{author}{\bibfnamefont{P.}~\bibnamefont{Kosec}},
  \bibinfo{author}{\bibfnamefont{V.}~\bibnamefont{Karas}},
  \bibinfo{author}{\bibfnamefont{M.}~\bibnamefont{Masterson}},
  \bibinfo{author}{\bibfnamefont{A.}~\bibnamefont{Mummery}},
  \bibnamefont{et~al.}, \bibinfo{journal}{Science Advances}
  \textbf{\bibinfo{volume}{10}}, \bibinfo{pages}{eadj8898}
  (\bibinfo{year}{2024}),
  \eprint{https://www.science.org/doi/pdf/10.1126/sciadv.adj8898},
  \urlprefix\url{https://www.science.org/doi/abs/10.1126/sciadv.adj8898}.

\bibitem[{\citenamefont{Dittmann and Miller}(2020)}]{Dittmann2020}
\bibinfo{author}{\bibfnamefont{A.~J.} \bibnamefont{Dittmann}} \bibnamefont{and}
  \bibinfo{author}{\bibfnamefont{M.~C.} \bibnamefont{Miller}},
  \bibinfo{journal}{Monthly Notices of the Royal Astronomical Society}
  \textbf{\bibinfo{volume}{493}}, \bibinfo{pages}{3732} (\bibinfo{year}{2020}),
  ISSN \bibinfo{issn}{0035-8711},
  \eprint{https://academic.oup.com/mnras/article-pdf/493/3/3732/32903023/staa463.pdf},
  \urlprefix\url{https://doi.org/10.1093/mnras/staa463}.

\bibitem[{\citenamefont{Levin}(2006)}]{Levin:2007}
\bibinfo{author}{\bibfnamefont{Y.}~\bibnamefont{Levin}},
  \bibinfo{journal}{Monthly Notices of the Royal Astronomical Society}
  \textbf{\bibinfo{volume}{374}}, \bibinfo{pages}{515} (\bibinfo{year}{2006}),
  ISSN \bibinfo{issn}{0035-8711},
  \eprint{https://academic.oup.com/mnras/article-pdf/374/2/515/4872435/mnras0374-0515.pdf},
  \urlprefix\url{https://doi.org/10.1111/j.1365-2966.2006.11155.x}.

\bibitem[{\citenamefont{Chakrabarti}(1996)}]{chakrabartiGravitationalWaveEmission1996}
\bibinfo{author}{\bibfnamefont{S.~K.} \bibnamefont{Chakrabarti}},
  \bibinfo{journal}{Physical Review D} \textbf{\bibinfo{volume}{53}},
  \bibinfo{pages}{2901} (\bibinfo{year}{1996}), ISSN \bibinfo{issn}{0556-2821,
  1089-4918}.

\bibitem[{\citenamefont{Shapiro and
  Marchant}(1978)}]{shapiroStarClustersContaining1978}
\bibinfo{author}{\bibfnamefont{S.~L.} \bibnamefont{Shapiro}} \bibnamefont{and}
  \bibinfo{author}{\bibfnamefont{A.~B.} \bibnamefont{Marchant}},
  \bibinfo{journal}{The Astrophysical Journal} \textbf{\bibinfo{volume}{225}},
  \bibinfo{pages}{603} (\bibinfo{year}{1978}), ISSN \bibinfo{issn}{0004-637X}.

\bibitem[{\citenamefont{{Bar-Or} and
  Alexander}(2016)}]{bar-orSTEADYSTATERELATIVISTICStelLAR2016}
\bibinfo{author}{\bibfnamefont{B.}~\bibnamefont{{Bar-Or}}} \bibnamefont{and}
  \bibinfo{author}{\bibfnamefont{T.}~\bibnamefont{Alexander}},
  \bibinfo{journal}{The Astrophysical Journal} \textbf{\bibinfo{volume}{820}},
  \bibinfo{pages}{129} (\bibinfo{year}{2016}), ISSN \bibinfo{issn}{0004-637X}.

\bibitem[{\citenamefont{Pan et~al.}(2021)\citenamefont{Pan, Lyu, and
  Yang}}]{Pan:2021oob}
\bibinfo{author}{\bibfnamefont{Z.}~\bibnamefont{Pan}},
  \bibinfo{author}{\bibfnamefont{Z.}~\bibnamefont{Lyu}}, \bibnamefont{and}
  \bibinfo{author}{\bibfnamefont{H.}~\bibnamefont{Yang}},
  \bibinfo{journal}{Phys. Rev. D} \textbf{\bibinfo{volume}{104}},
  \bibinfo{pages}{063007} (\bibinfo{year}{2021}), \eprint{2104.01208}.

\bibitem[{\citenamefont{Pan and Yang}(2021)}]{Pan:2021ksp}
\bibinfo{author}{\bibfnamefont{Z.}~\bibnamefont{Pan}} \bibnamefont{and}
  \bibinfo{author}{\bibfnamefont{H.}~\bibnamefont{Yang}},
  \bibinfo{journal}{Phys. Rev. D} \textbf{\bibinfo{volume}{103}},
  \bibinfo{pages}{103018} (\bibinfo{year}{2021}), \eprint{2101.09146}.

\bibitem[{\citenamefont{{Levin}}(2007)}]{Levin2007}
\bibinfo{author}{\bibfnamefont{Y.}~\bibnamefont{{Levin}}},
  \bibinfo{journal}{Monthly Notices of the Royal Astronomical Society}
  \textbf{\bibinfo{volume}{374}}, \bibinfo{pages}{515} (\bibinfo{year}{2007}),
  \eprint{astro-ph/0603583}.

\bibitem[{\citenamefont{{Goodman} and {Tan}}(2004)}]{GoodmanTan2004}
\bibinfo{author}{\bibfnamefont{J.}~\bibnamefont{{Goodman}}} \bibnamefont{and}
  \bibinfo{author}{\bibfnamefont{J.~C.} \bibnamefont{{Tan}}},
  \bibinfo{journal}{\apj} \textbf{\bibinfo{volume}{608}}, \bibinfo{pages}{108}
  (\bibinfo{year}{2004}), \eprint{astro-ph/0307361}.

\bibitem[{\citenamefont{Derdzinski and
  Mayer}(2023)}]{derdzinskiInsituExtremeMass2023}
\bibinfo{author}{\bibfnamefont{A.}~\bibnamefont{Derdzinski}} \bibnamefont{and}
  \bibinfo{author}{\bibfnamefont{L.}~\bibnamefont{Mayer}},
  \bibinfo{journal}{Monthly Notices of the Royal Astronomical Society}
  \textbf{\bibinfo{volume}{521}}, \bibinfo{pages}{4522} (\bibinfo{year}{2023}),
  ISSN \bibinfo{issn}{0035-8711, 1365-2966}, \eprint{2205.10382}.

\bibitem[{\citenamefont{{Zwick} et~al.}(2022)\citenamefont{{Zwick},
  {Derdzinski}, {Garg}, {Capelo}, and {Mayer}}}]{Zwick2022}
\bibinfo{author}{\bibfnamefont{L.}~\bibnamefont{{Zwick}}},
  \bibinfo{author}{\bibfnamefont{A.}~\bibnamefont{{Derdzinski}}},
  \bibinfo{author}{\bibfnamefont{M.}~\bibnamefont{{Garg}}},
  \bibinfo{author}{\bibfnamefont{P.~R.} \bibnamefont{{Capelo}}},
  \bibnamefont{and} \bibinfo{author}{\bibfnamefont{L.}~\bibnamefont{{Mayer}}},
  \bibinfo{journal}{Monthly Notices of the Royal Astronomical Society}
  \textbf{\bibinfo{volume}{511}}, \bibinfo{pages}{6143} (\bibinfo{year}{2022}),
  \eprint{2110.09097}.

\bibitem[{\citenamefont{Duque et~al.}(2025)\citenamefont{Duque, Kejriwal,
  Sberna, Speri, and Gair}}]{Duque:2024mfw}
\bibinfo{author}{\bibfnamefont{F.}~\bibnamefont{Duque}},
  \bibinfo{author}{\bibfnamefont{S.}~\bibnamefont{Kejriwal}},
  \bibinfo{author}{\bibfnamefont{L.}~\bibnamefont{Sberna}},
  \bibinfo{author}{\bibfnamefont{L.}~\bibnamefont{Speri}}, \bibnamefont{and}
  \bibinfo{author}{\bibfnamefont{J.}~\bibnamefont{Gair}},
  \bibinfo{journal}{Phys. Rev. D} \textbf{\bibinfo{volume}{111}},
  \bibinfo{pages}{084006} (\bibinfo{year}{2025}), \eprint{2411.03436}.

\bibitem[{\citenamefont{Kejriwal
  et~al.}(2024{\natexlab{b}})\citenamefont{Kejriwal, Speri, and
  Chua}}]{Kejriwal:2023djc}
\bibinfo{author}{\bibfnamefont{S.}~\bibnamefont{Kejriwal}},
  \bibinfo{author}{\bibfnamefont{L.}~\bibnamefont{Speri}}, \bibnamefont{and}
  \bibinfo{author}{\bibfnamefont{A.~J.~K.} \bibnamefont{Chua}},
  \bibinfo{journal}{Phys. Rev. D} \textbf{\bibinfo{volume}{110}},
  \bibinfo{pages}{084060} (\bibinfo{year}{2024}{\natexlab{b}}),
  \eprint{2312.13028}.

\bibitem[{\citenamefont{Khalvati et~al.}(2024)\citenamefont{Khalvati, Santini,
  Duque, Speri, Gair, Yang, and Brito}}]{Khalvati:2024tzz}
\bibinfo{author}{\bibfnamefont{H.}~\bibnamefont{Khalvati}},
  \bibinfo{author}{\bibfnamefont{A.}~\bibnamefont{Santini}},
  \bibinfo{author}{\bibfnamefont{F.}~\bibnamefont{Duque}},
  \bibinfo{author}{\bibfnamefont{L.}~\bibnamefont{Speri}},
  \bibinfo{author}{\bibfnamefont{J.}~\bibnamefont{Gair}},
  \bibinfo{author}{\bibfnamefont{H.}~\bibnamefont{Yang}}, \bibnamefont{and}
  \bibinfo{author}{\bibfnamefont{R.}~\bibnamefont{Brito}}
  (\bibinfo{year}{2024}), \eprint{2410.17310}.

\bibitem[{\citenamefont{Derdzinski et~al.}(2019)\citenamefont{Derdzinski,
  D'Orazio, Duffell, Haiman, and MacFadyen}}]{derdzinskiProbingGasDisc2019}
\bibinfo{author}{\bibfnamefont{A.~M.} \bibnamefont{Derdzinski}},
  \bibinfo{author}{\bibfnamefont{D.}~\bibnamefont{D'Orazio}},
  \bibinfo{author}{\bibfnamefont{P.}~\bibnamefont{Duffell}},
  \bibinfo{author}{\bibfnamefont{Z.}~\bibnamefont{Haiman}}, \bibnamefont{and}
  \bibinfo{author}{\bibfnamefont{A.}~\bibnamefont{MacFadyen}},
  \bibinfo{journal}{Monthly Notices of the Royal Astronomical Society}
  \textbf{\bibinfo{volume}{486}}, \bibinfo{pages}{2754} (\bibinfo{year}{2019}),
  ISSN \bibinfo{issn}{0035-8711, 1365-2966}.

\bibitem[{\citenamefont{Duffell et~al.}(2014)\citenamefont{Duffell, Haiman,
  MacFadyen, D'Orazio, and Farris}}]{duffellMigrationGapOpeningPlanets2014}
\bibinfo{author}{\bibfnamefont{P.~C.} \bibnamefont{Duffell}},
  \bibinfo{author}{\bibfnamefont{Z.}~\bibnamefont{Haiman}},
  \bibinfo{author}{\bibfnamefont{A.~I.} \bibnamefont{MacFadyen}},
  \bibinfo{author}{\bibfnamefont{D.~J.} \bibnamefont{D'Orazio}},
  \bibnamefont{and} \bibinfo{author}{\bibfnamefont{B.~D.}
  \bibnamefont{Farris}}, \bibinfo{journal}{The Astrophysical Journal}
  \textbf{\bibinfo{volume}{792}}, \bibinfo{pages}{L10} (\bibinfo{year}{2014}),
  ISSN \bibinfo{issn}{2041-8213}, \eprint{1405.3711}.

\bibitem[{\citenamefont{Wu et~al.}(2023)\citenamefont{Wu, Chen, and
  Lin}}]{Wu:2023qeh}
\bibinfo{author}{\bibfnamefont{Y.}~\bibnamefont{Wu}},
  \bibinfo{author}{\bibfnamefont{Y.-X.} \bibnamefont{Chen}}, \bibnamefont{and}
  \bibinfo{author}{\bibfnamefont{D.~N.~C.} \bibnamefont{Lin}},
  \bibinfo{journal}{Mon. Not. Roy. Astron. Soc.}
  \textbf{\bibinfo{volume}{528}}, \bibinfo{pages}{L127} (\bibinfo{year}{2023}),
  \eprint{2311.15747}.

\bibitem[{\citenamefont{Derdzinski et~al.}(2021)\citenamefont{Derdzinski,
  D'Orazio, Duffell, Haiman, and MacFadyen}}]{derdzinskiEvolutionGasDisc2021}
\bibinfo{author}{\bibfnamefont{A.}~\bibnamefont{Derdzinski}},
  \bibinfo{author}{\bibfnamefont{D.}~\bibnamefont{D'Orazio}},
  \bibinfo{author}{\bibfnamefont{P.}~\bibnamefont{Duffell}},
  \bibinfo{author}{\bibfnamefont{Z.}~\bibnamefont{Haiman}}, \bibnamefont{and}
  \bibinfo{author}{\bibfnamefont{A.}~\bibnamefont{MacFadyen}},
  \bibinfo{journal}{Monthly Notices of the Royal Astronomical Society}
  \textbf{\bibinfo{volume}{501}}, \bibinfo{pages}{3540} (\bibinfo{year}{2021}),
  ISSN \bibinfo{issn}{0035-8711, 1365-2966}.

\bibitem[{\citenamefont{Goldreich and Tremaine}(1979)}]{Goldreich1979}
\bibinfo{author}{\bibfnamefont{P.}~\bibnamefont{Goldreich}} \bibnamefont{and}
  \bibinfo{author}{\bibfnamefont{S.}~\bibnamefont{Tremaine}},
  \bibinfo{journal}{The Astrophysical Journal} \textbf{\bibinfo{volume}{233}},
  \bibinfo{pages}{857} (\bibinfo{year}{1979}), ISSN \bibinfo{issn}{1538-4357},
  \urlprefix\url{http://dx.doi.org/10.1086/157448}.

\bibitem[{\citenamefont{Goldreich and
  Tremaine}(1980)}]{goldreichDisksatelliteInteractions1980}
\bibinfo{author}{\bibfnamefont{P.}~\bibnamefont{Goldreich}} \bibnamefont{and}
  \bibinfo{author}{\bibfnamefont{S.}~\bibnamefont{Tremaine}},
  \bibinfo{journal}{The Astrophysical Journal} \textbf{\bibinfo{volume}{241}},
  \bibinfo{pages}{425} (\bibinfo{year}{1980}), ISSN \bibinfo{issn}{0004-637X}.

\bibitem[{\citenamefont{Baruteau and
  Masset}(2013)}]{baruteauRecentDevelopmentsPlanet2013}
\bibinfo{author}{\bibfnamefont{C.}~\bibnamefont{Baruteau}} \bibnamefont{and}
  \bibinfo{author}{\bibfnamefont{F.}~\bibnamefont{Masset}},
  \emph{\bibinfo{title}{Recent Developments in Planet Migration Theory}}
  (\bibinfo{publisher}{Springer Berlin Heidelberg}, \bibinfo{address}{Berlin,
  Heidelberg}, \bibinfo{year}{2013}), pp. \bibinfo{pages}{201--253}, ISBN
  \bibinfo{isbn}{978-3-642-32961-6},
  \urlprefix\url{https://doi.org/10.1007/978-3-642-32961-6_6}.

\bibitem[{\citenamefont{Nelson}(2018)}]{nelsonPlanetaryMigrationProtoplanetary2018}
\bibinfo{author}{\bibfnamefont{R.~P.} \bibnamefont{Nelson}}, in
  \emph{\bibinfo{booktitle}{Handbook of {{Exoplanets}}}}, edited by
  \bibinfo{editor}{\bibfnamefont{H.~J.} \bibnamefont{Deeg}} \bibnamefont{and}
  \bibinfo{editor}{\bibfnamefont{J.~A.} \bibnamefont{Belmonte}}
  (\bibinfo{publisher}{Springer International Publishing},
  \bibinfo{address}{Cham}, \bibinfo{year}{2018}), pp.
  \bibinfo{pages}{2287--2317}, ISBN \bibinfo{isbn}{978-3-319-55332-0
  978-3-319-55333-7}.

\bibitem[{\citenamefont{Baruteau et~al.}(2014)\citenamefont{Baruteau, Crida,
  Paardekooper, Masset, Guilet, Bitsch, Nelson, Kley, and
  Papaloizou}}]{baruteauPlanetDiscInteractionsEarly2014}
\bibinfo{author}{\bibfnamefont{C.}~\bibnamefont{Baruteau}},
  \bibinfo{author}{\bibfnamefont{A.}~\bibnamefont{Crida}},
  \bibinfo{author}{\bibfnamefont{S.-J.} \bibnamefont{Paardekooper}},
  \bibinfo{author}{\bibfnamefont{F.}~\bibnamefont{Masset}},
  \bibinfo{author}{\bibfnamefont{J.}~\bibnamefont{Guilet}},
  \bibinfo{author}{\bibfnamefont{B.}~\bibnamefont{Bitsch}},
  \bibinfo{author}{\bibfnamefont{R.~P.} \bibnamefont{Nelson}},
  \bibinfo{author}{\bibfnamefont{W.}~\bibnamefont{Kley}}, \bibnamefont{and}
  \bibinfo{author}{\bibfnamefont{J.~C.~B.} \bibnamefont{Papaloizou}}, in
  \emph{\bibinfo{booktitle}{Protostars and Planets VI}}
  (\bibinfo{publisher}{University of Arizona Press}, \bibinfo{year}{2014}),
  \eprint{1312.4293}.

\bibitem[{\citenamefont{Tanaka and
  Ward}(2004)}]{tanakaThreeDimensionalInteraction2004}
\bibinfo{author}{\bibfnamefont{H.}~\bibnamefont{Tanaka}} \bibnamefont{and}
  \bibinfo{author}{\bibfnamefont{W.~R.} \bibnamefont{Ward}},
  \bibinfo{journal}{The Astrophysical Journal} \textbf{\bibinfo{volume}{602}},
  \bibinfo{pages}{388} (\bibinfo{year}{2004}), ISSN \bibinfo{issn}{0004-637X,
  1538-4357}.

\bibitem[{\citenamefont{Tanaka et~al.}(2002)\citenamefont{Tanaka, Takeuchi, and
  Ward}}]{tanakaThreeDimensionalInteraction2002}
\bibinfo{author}{\bibfnamefont{H.}~\bibnamefont{Tanaka}},
  \bibinfo{author}{\bibfnamefont{T.}~\bibnamefont{Takeuchi}}, \bibnamefont{and}
  \bibinfo{author}{\bibfnamefont{W.~R.} \bibnamefont{Ward}},
  \bibinfo{journal}{The Astrophysical Journal} \textbf{\bibinfo{volume}{565}},
  \bibinfo{pages}{1257} (\bibinfo{year}{2002}), ISSN \bibinfo{issn}{0004-637X,
  1538-4357}.

\bibitem[{\citenamefont{Taracchini et~al.}(2014)\citenamefont{Taracchini,
  Buonanno, Khanna, and Hughes}}]{PhysRevD.90.084025}
\bibinfo{author}{\bibfnamefont{A.}~\bibnamefont{Taracchini}},
  \bibinfo{author}{\bibfnamefont{A.}~\bibnamefont{Buonanno}},
  \bibinfo{author}{\bibfnamefont{G.}~\bibnamefont{Khanna}}, \bibnamefont{and}
  \bibinfo{author}{\bibfnamefont{S.~A.} \bibnamefont{Hughes}},
  \bibinfo{journal}{Phys. Rev. D} \textbf{\bibinfo{volume}{90}},
  \bibinfo{pages}{084025} (\bibinfo{year}{2014}),
  \urlprefix\url{https://link.aps.org/doi/10.1103/PhysRevD.90.084025}.

\bibitem[{\citenamefont{Tanaka and Okada}(2024)}]{Tanaka_2024}
\bibinfo{author}{\bibfnamefont{H.}~\bibnamefont{Tanaka}} \bibnamefont{and}
  \bibinfo{author}{\bibfnamefont{K.}~\bibnamefont{Okada}},
  \bibinfo{journal}{The Astrophysical Journal} \textbf{\bibinfo{volume}{968}},
  \bibinfo{pages}{28} (\bibinfo{year}{2024}),
  \urlprefix\url{https://dx.doi.org/10.3847/1538-4357/ad410d}.

\bibitem[{\citenamefont{Korycansky and
  Pollack}(1993)}]{korycanskyNumericalCalculationsLinear1993}
\bibinfo{author}{\bibfnamefont{D.~G.} \bibnamefont{Korycansky}}
  \bibnamefont{and} \bibinfo{author}{\bibfnamefont{J.~B.}
  \bibnamefont{Pollack}}, \bibinfo{journal}{Icarus}
  \textbf{\bibinfo{volume}{102}}, \bibinfo{pages}{150} (\bibinfo{year}{1993}),
  ISSN \bibinfo{issn}{0019-1035}.

\bibitem[{\citenamefont{Lin and Papaloizou}(1993)}]{Levy1993ProtostarsAP}
\bibinfo{author}{\bibfnamefont{D.~N.~C.} \bibnamefont{Lin}} \bibnamefont{and}
  \bibinfo{author}{\bibfnamefont{J.~C.~B.} \bibnamefont{Papaloizou}}, in
  \emph{\bibinfo{booktitle}{Protostars and planets III}}, edited by
  \bibinfo{editor}{\bibfnamefont{E.~H.} \bibnamefont{Levy}} \bibnamefont{and}
  \bibinfo{editor}{\bibfnamefont{J.~I.} \bibnamefont{Lunine}}
  (\bibinfo{publisher}{Arizona University Press}, \bibinfo{year}{1993}),
  \urlprefix\url{https://uapress.arizona.edu/book/protostars-and-planets-iii}.

\bibitem[{\citenamefont{Lin and Papaloizou}(1986)}]{Lin1986}
\bibinfo{author}{\bibfnamefont{D.~N.~C.} \bibnamefont{Lin}} \bibnamefont{and}
  \bibinfo{author}{\bibfnamefont{J.}~\bibnamefont{Papaloizou}},
  \bibinfo{journal}{The Astrophysical Journal} \textbf{\bibinfo{volume}{307}},
  \bibinfo{pages}{395} (\bibinfo{year}{1986}), ISSN \bibinfo{issn}{1538-4357},
  \urlprefix\url{http://dx.doi.org/10.1086/164426}.

\bibitem[{\citenamefont{Malik et~al.}(2015)\citenamefont{Malik, Meru, Mayer,
  and Meyer}}]{Malik2015}
\bibinfo{author}{\bibfnamefont{M.}~\bibnamefont{Malik}},
  \bibinfo{author}{\bibfnamefont{F.}~\bibnamefont{Meru}},
  \bibinfo{author}{\bibfnamefont{L.}~\bibnamefont{Mayer}}, \bibnamefont{and}
  \bibinfo{author}{\bibfnamefont{M.}~\bibnamefont{Meyer}},
  \bibinfo{journal}{The Astrophysical Journal} \textbf{\bibinfo{volume}{802}},
  \bibinfo{pages}{56} (\bibinfo{year}{2015}), ISSN \bibinfo{issn}{1538-4357},
  \urlprefix\url{http://dx.doi.org/10.1088/0004-637X/802/1/56}.

\bibitem[{\citenamefont{Crida et~al.}(2006)\citenamefont{Crida, Morbidelli, and
  Masset}}]{cridaWidthShapeGaps2006}
\bibinfo{author}{\bibfnamefont{A.}~\bibnamefont{Crida}},
  \bibinfo{author}{\bibfnamefont{A.}~\bibnamefont{Morbidelli}},
  \bibnamefont{and} \bibinfo{author}{\bibfnamefont{F.}~\bibnamefont{Masset}},
  \bibinfo{journal}{Icarus} \textbf{\bibinfo{volume}{181}},
  \bibinfo{pages}{587} (\bibinfo{year}{2006}), ISSN \bibinfo{issn}{0019-1035}.

\bibitem[{\citenamefont{Ward and Hourigan}(1989)}]{Ward1989}
\bibinfo{author}{\bibfnamefont{W.~R.} \bibnamefont{Ward}} \bibnamefont{and}
  \bibinfo{author}{\bibfnamefont{K.}~\bibnamefont{Hourigan}},
  \bibinfo{journal}{The Astrophysical Journal} \textbf{\bibinfo{volume}{347}},
  \bibinfo{pages}{490} (\bibinfo{year}{1989}), ISSN \bibinfo{issn}{1538-4357},
  \urlprefix\url{http://dx.doi.org/10.1086/168138}.

\bibitem[{\citenamefont{Garg et~al.}(2022)\citenamefont{Garg, Derdzinski,
  Zwick, Capelo, and Mayer}}]{gargImprintGasGravitational2022a}
\bibinfo{author}{\bibfnamefont{M.}~\bibnamefont{Garg}},
  \bibinfo{author}{\bibfnamefont{A.}~\bibnamefont{Derdzinski}},
  \bibinfo{author}{\bibfnamefont{L.}~\bibnamefont{Zwick}},
  \bibinfo{author}{\bibfnamefont{P.~R.} \bibnamefont{Capelo}},
  \bibnamefont{and} \bibinfo{author}{\bibfnamefont{L.}~\bibnamefont{Mayer}},
  \bibinfo{journal}{Monthly Notices of the Royal Astronomical Society}
  \textbf{\bibinfo{volume}{517}}, \bibinfo{pages}{1339} (\bibinfo{year}{2022}),
  ISSN \bibinfo{issn}{0035-8711, 1365-2966}, \eprint{2206.05292}.

\bibitem[{\citenamefont{Nelson et~al.}(2000)\citenamefont{Nelson, Papaloizou,
  Masset, and Kley}}]{nelsonMigrationGrowthProtoplanets2000}
\bibinfo{author}{\bibfnamefont{R.~P.} \bibnamefont{Nelson}},
  \bibinfo{author}{\bibfnamefont{J.~C.~B.} \bibnamefont{Papaloizou}},
  \bibinfo{author}{\bibfnamefont{F.}~\bibnamefont{Masset}}, \bibnamefont{and}
  \bibinfo{author}{\bibfnamefont{W.}~\bibnamefont{Kley}},
  \bibinfo{journal}{Monthly Notices of the Royal Astronomical Society}
  \textbf{\bibinfo{volume}{318}}, \bibinfo{pages}{18} (\bibinfo{year}{2000}),
  ISSN \bibinfo{issn}{0035-8711, 1365-2966}.

\bibitem[{\citenamefont{Duffell and MacFadyen}(2013)}]{Duffell2013}
\bibinfo{author}{\bibfnamefont{P.~C.} \bibnamefont{Duffell}} \bibnamefont{and}
  \bibinfo{author}{\bibfnamefont{A.~I.} \bibnamefont{MacFadyen}},
  \bibinfo{journal}{The Astrophysical Journal} \textbf{\bibinfo{volume}{769}},
  \bibinfo{pages}{41} (\bibinfo{year}{2013}), ISSN \bibinfo{issn}{1538-4357},
  \urlprefix\url{http://dx.doi.org/10.1088/0004-637X/769/1/41}.

\bibitem[{\citenamefont{Kanagawa et~al.}(2018)\citenamefont{Kanagawa, Tanaka,
  and Szuszkiewicz}}]{kanagawaRadialMigrationGapopening2018}
\bibinfo{author}{\bibfnamefont{K.~D.} \bibnamefont{Kanagawa}},
  \bibinfo{author}{\bibfnamefont{H.}~\bibnamefont{Tanaka}}, \bibnamefont{and}
  \bibinfo{author}{\bibfnamefont{E.}~\bibnamefont{Szuszkiewicz}},
  \bibinfo{journal}{The Astrophysical Journal} \textbf{\bibinfo{volume}{861}},
  \bibinfo{pages}{140} (\bibinfo{year}{2018}), ISSN \bibinfo{issn}{0004-637X,
  1538-4357}, \eprint{1805.11101}.

\bibitem[{\citenamefont{Nelson}(2005)}]{nelsonOrbitalEvolutionLow2005}
\bibinfo{author}{\bibfnamefont{R.~P.} \bibnamefont{Nelson}},
  \bibinfo{journal}{Astronomy \& Astrophysics} \textbf{\bibinfo{volume}{443}},
  \bibinfo{pages}{1067} (\bibinfo{year}{2005}), ISSN \bibinfo{issn}{0004-6361,
  1432-0746}.

\bibitem[{\citenamefont{Baruteau and Lin}(2010)}]{Baruteau_2010}
\bibinfo{author}{\bibfnamefont{C.}~\bibnamefont{Baruteau}} \bibnamefont{and}
  \bibinfo{author}{\bibfnamefont{D.~N.~C.} \bibnamefont{Lin}},
  \bibinfo{journal}{The Astrophysical Journal} \textbf{\bibinfo{volume}{709}},
  \bibinfo{pages}{759} (\bibinfo{year}{2010}),
  \urlprefix\url{https://dx.doi.org/10.1088/0004-637X/709/2/759}.

\bibitem[{\citenamefont{{Wu} et~al.}(2024)\citenamefont{{Wu}, {Chen}, and
  {Lin}}}]{WuChenLin2024}
\bibinfo{author}{\bibfnamefont{Y.}~\bibnamefont{{Wu}}},
  \bibinfo{author}{\bibfnamefont{Y.-X.} \bibnamefont{{Chen}}},
  \bibnamefont{and} \bibinfo{author}{\bibfnamefont{D.~N.~C.}
  \bibnamefont{{Lin}}}, \bibinfo{journal}{Monthly Notices of the Royal
  Astronomical Society} \textbf{\bibinfo{volume}{528}}, \bibinfo{pages}{L127}
  (\bibinfo{year}{2024}), \eprint{2311.15747}.

\bibitem[{\citenamefont{Aoyama and Bai}(2023)}]{Aoyama_2023}
\bibinfo{author}{\bibfnamefont{Y.}~\bibnamefont{Aoyama}} \bibnamefont{and}
  \bibinfo{author}{\bibfnamefont{X.-N.} \bibnamefont{Bai}},
  \bibinfo{journal}{The Astrophysical Journal} \textbf{\bibinfo{volume}{946}},
  \bibinfo{pages}{5} (\bibinfo{year}{2023}),
  \urlprefix\url{https://dx.doi.org/10.3847/1538-4357/acb81f}.

\bibitem[{\citenamefont{Frank et~al.}(2002)\citenamefont{Frank, King, and
  Raine}}]{frankAccretionPowerAstrophysics2002}
\bibinfo{author}{\bibfnamefont{J.}~\bibnamefont{Frank}},
  \bibinfo{author}{\bibfnamefont{A.~R.} \bibnamefont{King}}, \bibnamefont{and}
  \bibinfo{author}{\bibfnamefont{D.}~\bibnamefont{Raine}},
  \emph{\bibinfo{title}{Accretion Power in Astrophysics}}
  (\bibinfo{publisher}{Cambridge Univ. Press}, \bibinfo{address}{Cambridge},
  \bibinfo{year}{2002}), \bibinfo{edition}{3rd} ed., ISBN
  \bibinfo{isbn}{978-0-521-62957-7 978-0-521-62053-6}.

\bibitem[{\citenamefont{{Masset}}(2002)}]{Masset2002}
\bibinfo{author}{\bibfnamefont{F.~S.} \bibnamefont{{Masset}}},
  \bibinfo{journal}{Astronomy and Astrophysics} \textbf{\bibinfo{volume}{387}},
  \bibinfo{pages}{605} (\bibinfo{year}{2002}), \eprint{astro-ph/0205211}.

\bibitem[{\citenamefont{Teukolsky}(1973)}]{Teukolsky:1973ha}
\bibinfo{author}{\bibfnamefont{S.~A.} \bibnamefont{Teukolsky}},
  \bibinfo{journal}{Astrophys. J.} \textbf{\bibinfo{volume}{185}},
  \bibinfo{pages}{635} (\bibinfo{year}{1973}).

\bibitem[{\citenamefont{Drasco and
  Hughes}(2006)}]{drascoGravitationalWaveSnapshots2006}
\bibinfo{author}{\bibfnamefont{S.}~\bibnamefont{Drasco}} \bibnamefont{and}
  \bibinfo{author}{\bibfnamefont{S.~A.} \bibnamefont{Hughes}},
  \bibinfo{journal}{Physical Review D} \textbf{\bibinfo{volume}{73}},
  \bibinfo{pages}{024027} (\bibinfo{year}{2006}), ISSN
  \bibinfo{issn}{1550-7998, 1550-2368}.

\bibitem[{\citenamefont{Bardeen et~al.}(1972)\citenamefont{Bardeen, Press, and
  Teukolsky}}]{bardeenRotatingBlackHoles1972}
\bibinfo{author}{\bibfnamefont{J.~M.} \bibnamefont{Bardeen}},
  \bibinfo{author}{\bibfnamefont{W.~H.} \bibnamefont{Press}}, \bibnamefont{and}
  \bibinfo{author}{\bibfnamefont{S.~A.} \bibnamefont{Teukolsky}},
  \bibinfo{journal}{The Astrophysical Journal} \textbf{\bibinfo{volume}{178}},
  \bibinfo{pages}{347} (\bibinfo{year}{1972}), ISSN \bibinfo{issn}{0004-637X,
  1538-4357}.

\bibitem[{\citenamefont{Chua et~al.}(2021)\citenamefont{Chua, Katz, Warburton,
  and Hughes}}]{Chua:2020stf}
\bibinfo{author}{\bibfnamefont{A.~J.~K.} \bibnamefont{Chua}},
  \bibinfo{author}{\bibfnamefont{M.~L.} \bibnamefont{Katz}},
  \bibinfo{author}{\bibfnamefont{N.}~\bibnamefont{Warburton}},
  \bibnamefont{and} \bibinfo{author}{\bibfnamefont{S.~A.}
  \bibnamefont{Hughes}}, \emph{\bibinfo{title}{{Rapid generation of fully
  relativistic extreme-mass-ratio-inspiral waveform templates for LISA data
  analysis}}} (\bibinfo{year}{2021}), \eprint{2008.06071}.

\bibitem[{\citenamefont{Katz et~al.}(2021)\citenamefont{Katz, Chua, Speri,
  Warburton, and Hughes}}]{katzFastEMRIWaveformsNewTools2021}
\bibinfo{author}{\bibfnamefont{M.~L.} \bibnamefont{Katz}},
  \bibinfo{author}{\bibfnamefont{A.~J.~K.} \bibnamefont{Chua}},
  \bibinfo{author}{\bibfnamefont{L.}~\bibnamefont{Speri}},
  \bibinfo{author}{\bibfnamefont{N.}~\bibnamefont{Warburton}},
  \bibnamefont{and} \bibinfo{author}{\bibfnamefont{S.~A.}
  \bibnamefont{Hughes}}, \bibinfo{journal}{Physical Review D}
  \textbf{\bibinfo{volume}{104}}, \bibinfo{pages}{064047}
  (\bibinfo{year}{2021}), ISSN \bibinfo{issn}{2470-0010, 2470-0029},
  \eprint{2104.04582}.

\bibitem[{\citenamefont{Speri et~al.}(2023{\natexlab{b}})\citenamefont{Speri,
  Katz, Chua, Hughes, Warburton, Thompson, Chapman-Bird, and
  Gair}}]{Speri:2023jte}
\bibinfo{author}{\bibfnamefont{L.}~\bibnamefont{Speri}},
  \bibinfo{author}{\bibfnamefont{M.~L.} \bibnamefont{Katz}},
  \bibinfo{author}{\bibfnamefont{A.~J.~K.} \bibnamefont{Chua}},
  \bibinfo{author}{\bibfnamefont{S.~A.} \bibnamefont{Hughes}},
  \bibinfo{author}{\bibfnamefont{N.}~\bibnamefont{Warburton}},
  \bibinfo{author}{\bibfnamefont{J.~E.} \bibnamefont{Thompson}},
  \bibinfo{author}{\bibfnamefont{C.~E.~A.} \bibnamefont{Chapman-Bird}},
  \bibnamefont{and} \bibinfo{author}{\bibfnamefont{J.~R.} \bibnamefont{Gair}},
  \emph{\bibinfo{title}{{Fast and Fourier: Extreme Mass Ratio Inspiral
  Waveforms in the Frequency Domain}}} (\bibinfo{year}{2023}{\natexlab{b}}),
  \eprint{2307.12585}.

\bibitem[{\citenamefont{Chua et~al.}(2017)\citenamefont{Chua, Moore, and
  Gair}}]{PhysRevD.96.044005}
\bibinfo{author}{\bibfnamefont{A.~J.~K.} \bibnamefont{Chua}},
  \bibinfo{author}{\bibfnamefont{C.~J.} \bibnamefont{Moore}}, \bibnamefont{and}
  \bibinfo{author}{\bibfnamefont{J.~R.} \bibnamefont{Gair}},
  \bibinfo{journal}{Phys. Rev. D} \textbf{\bibinfo{volume}{96}},
  \bibinfo{pages}{044005} (\bibinfo{year}{2017}),
  \urlprefix\url{https://link.aps.org/doi/10.1103/PhysRevD.96.044005}.

\bibitem[{\citenamefont{Barack and
  Cutler}(2004)}]{barackLISACaptureSources2004}
\bibinfo{author}{\bibfnamefont{L.}~\bibnamefont{Barack}} \bibnamefont{and}
  \bibinfo{author}{\bibfnamefont{C.}~\bibnamefont{Cutler}},
  \bibinfo{journal}{Physical Review D} \textbf{\bibinfo{volume}{69}},
  \bibinfo{pages}{082005} (\bibinfo{year}{2004}).

\bibitem[{\citenamefont{Peters and Mathews}(1963)}]{peters_gravitational_1963}
\bibinfo{author}{\bibfnamefont{P.~C.} \bibnamefont{Peters}} \bibnamefont{and}
  \bibinfo{author}{\bibfnamefont{J.}~\bibnamefont{Mathews}},
  \bibinfo{journal}{Physical Review} \textbf{\bibinfo{volume}{131}},
  \bibinfo{pages}{435} (\bibinfo{year}{1963}), ISSN \bibinfo{issn}{0031-899X},
  \urlprefix\url{https://link.aps.org/doi/10.1103/PhysRev.131.435}.

\bibitem[{\citenamefont{Zwick et~al.}(2022)\citenamefont{Zwick, Derdzinski,
  Garg, Capelo, and Mayer}}]{zwickDirtyWaveformsMultiband2022}
\bibinfo{author}{\bibfnamefont{L.}~\bibnamefont{Zwick}},
  \bibinfo{author}{\bibfnamefont{A.}~\bibnamefont{Derdzinski}},
  \bibinfo{author}{\bibfnamefont{M.}~\bibnamefont{Garg}},
  \bibinfo{author}{\bibfnamefont{P.~R.} \bibnamefont{Capelo}},
  \bibnamefont{and} \bibinfo{author}{\bibfnamefont{L.}~\bibnamefont{Mayer}},
  \bibinfo{journal}{Monthly Notices of the Royal Astronomical Society}
  \textbf{\bibinfo{volume}{511}}, \bibinfo{pages}{6143} (\bibinfo{year}{2022}),
  ISSN \bibinfo{issn}{0035-8711, 1365-2966}.

\bibitem[{\citenamefont{Karnesis et~al.}(2023)\citenamefont{Karnesis, Katz,
  Korsakova, Gair, and Stergioulas}}]{karnesisErynMultipurposeSampler2023}
\bibinfo{author}{\bibfnamefont{N.}~\bibnamefont{Karnesis}},
  \bibinfo{author}{\bibfnamefont{M.~L.} \bibnamefont{Katz}},
  \bibinfo{author}{\bibfnamefont{N.}~\bibnamefont{Korsakova}},
  \bibinfo{author}{\bibfnamefont{J.~R.} \bibnamefont{Gair}}, \bibnamefont{and}
  \bibinfo{author}{\bibfnamefont{N.}~\bibnamefont{Stergioulas}},
  \bibinfo{journal}{Monthly Notices of the Royal Astronomical Society}
  \textbf{\bibinfo{volume}{526}}, \bibinfo{pages}{4814} (\bibinfo{year}{2023}),
  ISSN \bibinfo{issn}{0035-8711, 1365-2966}, \eprint{2303.02164}.

\bibitem[{\citenamefont{Cutler}(1998)}]{cutlerAngularResolutionLISA1998}
\bibinfo{author}{\bibfnamefont{C.}~\bibnamefont{Cutler}},
  \bibinfo{journal}{Physical Review D} \textbf{\bibinfo{volume}{57}},
  \bibinfo{pages}{7089} (\bibinfo{year}{1998}), ISSN \bibinfo{issn}{0556-2821,
  1089-4918}.

\bibitem[{\citenamefont{Cutler and
  Flanagan}(1994)}]{cutlerGravitationalWavesMerging1994}
\bibinfo{author}{\bibfnamefont{C.}~\bibnamefont{Cutler}} \bibnamefont{and}
  \bibinfo{author}{\bibfnamefont{{\'E}.~E.} \bibnamefont{Flanagan}},
  \bibinfo{journal}{Physical Review D} \textbf{\bibinfo{volume}{49}},
  \bibinfo{pages}{2658} (\bibinfo{year}{1994}), ISSN \bibinfo{issn}{0556-2821}.

\bibitem[{\citenamefont{Team}()}]{SciRd}
\bibinfo{author}{\bibfnamefont{L.~S.~S.} \bibnamefont{Team}},
  \emph{\bibinfo{title}{Lisa science requirements document}},
  \urlprefix\url{https://www.cosmos.esa.int/documents/678316/1700384/SciRD.pdf}.

\bibitem[{\citenamefont{Strub et~al.}(2024)\citenamefont{Strub, Ferraioli,
  Schmelzbach, St\"ahler, and Giardini}}]{Strub:2024kbe}
\bibinfo{author}{\bibfnamefont{S.~H.} \bibnamefont{Strub}},
  \bibinfo{author}{\bibfnamefont{L.}~\bibnamefont{Ferraioli}},
  \bibinfo{author}{\bibfnamefont{C.}~\bibnamefont{Schmelzbach}},
  \bibinfo{author}{\bibfnamefont{S.~C.} \bibnamefont{St\"ahler}},
  \bibnamefont{and} \bibinfo{author}{\bibfnamefont{D.}~\bibnamefont{Giardini}},
  \bibinfo{journal}{Phys. Rev. D} \textbf{\bibinfo{volume}{110}},
  \bibinfo{pages}{024005} (\bibinfo{year}{2024}), \eprint{2403.15318}.

\bibitem[{\citenamefont{Littenberg and Cornish}(2023)}]{Littenberg:2023xpl}
\bibinfo{author}{\bibfnamefont{T.~B.} \bibnamefont{Littenberg}}
  \bibnamefont{and} \bibinfo{author}{\bibfnamefont{N.~J.}
  \bibnamefont{Cornish}}, \bibinfo{journal}{Phys. Rev. D}
  \textbf{\bibinfo{volume}{107}}, \bibinfo{pages}{063004}
  (\bibinfo{year}{2023}), \eprint{2301.03673}.

\bibitem[{\citenamefont{Katz et~al.}(2025)\citenamefont{Katz, Karnesis,
  Korsakova, Gair, and Stergioulas}}]{Katz:2024oqg}
\bibinfo{author}{\bibfnamefont{M.~L.} \bibnamefont{Katz}},
  \bibinfo{author}{\bibfnamefont{N.}~\bibnamefont{Karnesis}},
  \bibinfo{author}{\bibfnamefont{N.}~\bibnamefont{Korsakova}},
  \bibinfo{author}{\bibfnamefont{J.~R.} \bibnamefont{Gair}}, \bibnamefont{and}
  \bibinfo{author}{\bibfnamefont{N.}~\bibnamefont{Stergioulas}},
  \bibinfo{journal}{Phys. Rev. D} \textbf{\bibinfo{volume}{111}},
  \bibinfo{pages}{024060} (\bibinfo{year}{2025}), \eprint{2405.04690}.

\bibitem[{\citenamefont{Deng et~al.}(2025)\citenamefont{Deng, Babak, Le~Jeune,
  Marsat, Plagnol, and Sartirana}}]{Deng:2025wgk}
\bibinfo{author}{\bibfnamefont{S.}~\bibnamefont{Deng}},
  \bibinfo{author}{\bibfnamefont{S.}~\bibnamefont{Babak}},
  \bibinfo{author}{\bibfnamefont{M.}~\bibnamefont{Le~Jeune}},
  \bibinfo{author}{\bibfnamefont{S.}~\bibnamefont{Marsat}},
  \bibinfo{author}{\bibfnamefont{E.}~\bibnamefont{Plagnol}}, \bibnamefont{and}
  \bibinfo{author}{\bibfnamefont{A.}~\bibnamefont{Sartirana}},
  \emph{\bibinfo{title}{{Modular global-fit pipeline for LISA data analysis}}}
  (\bibinfo{year}{2025}), \eprint{2501.10277}.

\bibitem[{\citenamefont{{Zwick} et~al.}(2024)\citenamefont{{Zwick}, {Tiede},
  {Trani}, {Derdzinski}, {Haiman}, {D'Orazio}, and {Samsing}}}]{Zwick2024}
\bibinfo{author}{\bibfnamefont{L.}~\bibnamefont{{Zwick}}},
  \bibinfo{author}{\bibfnamefont{C.}~\bibnamefont{{Tiede}}},
  \bibinfo{author}{\bibfnamefont{A.~A.} \bibnamefont{{Trani}}},
  \bibinfo{author}{\bibfnamefont{A.}~\bibnamefont{{Derdzinski}}},
  \bibinfo{author}{\bibfnamefont{Z.}~\bibnamefont{{Haiman}}},
  \bibinfo{author}{\bibfnamefont{D.~J.} \bibnamefont{{D'Orazio}}},
  \bibnamefont{and}
  \bibinfo{author}{\bibfnamefont{J.}~\bibnamefont{{Samsing}}},
  \bibinfo{journal}{\prd} \textbf{\bibinfo{volume}{110}}, \bibinfo{eid}{103005}
  (\bibinfo{year}{2024}), \eprint{2405.05698}.

\bibitem[{\citenamefont{{Ennoggi} et~al.}(2025)\citenamefont{{Ennoggi},
  {Campanelli}, {Zlochower}, {Noble}, {Krolik}, {Cattorini}, {Kalinani},
  {Mewes}, {Chabanov}, {Ji} et~al.}}]{Ennoggi2025}
\bibinfo{author}{\bibfnamefont{L.}~\bibnamefont{{Ennoggi}}},
  \bibinfo{author}{\bibfnamefont{M.}~\bibnamefont{{Campanelli}}},
  \bibinfo{author}{\bibfnamefont{Y.}~\bibnamefont{{Zlochower}}},
  \bibinfo{author}{\bibfnamefont{S.~C.} \bibnamefont{{Noble}}},
  \bibinfo{author}{\bibfnamefont{J.}~\bibnamefont{{Krolik}}},
  \bibinfo{author}{\bibfnamefont{F.}~\bibnamefont{{Cattorini}}},
  \bibinfo{author}{\bibfnamefont{J.~V.} \bibnamefont{{Kalinani}}},
  \bibinfo{author}{\bibfnamefont{V.}~\bibnamefont{{Mewes}}},
  \bibinfo{author}{\bibfnamefont{M.}~\bibnamefont{{Chabanov}}},
  \bibinfo{author}{\bibfnamefont{L.}~\bibnamefont{{Ji}}}, \bibnamefont{et~al.},
  \bibinfo{journal}{arXiv e-prints} \bibinfo{eid}{arXiv:2502.06389}
  (\bibinfo{year}{2025}), \eprint{2502.06389}.

\bibitem[{\citenamefont{{Avara} et~al.}(2024)\citenamefont{{Avara}, {Krolik},
  {Campanelli}, {Noble}, {Bowen}, and {Ryu}}}]{Avara2024}
\bibinfo{author}{\bibfnamefont{M.~J.} \bibnamefont{{Avara}}},
  \bibinfo{author}{\bibfnamefont{J.~H.} \bibnamefont{{Krolik}}},
  \bibinfo{author}{\bibfnamefont{M.}~\bibnamefont{{Campanelli}}},
  \bibinfo{author}{\bibfnamefont{S.~C.} \bibnamefont{{Noble}}},
  \bibinfo{author}{\bibfnamefont{D.}~\bibnamefont{{Bowen}}}, \bibnamefont{and}
  \bibinfo{author}{\bibfnamefont{T.}~\bibnamefont{{Ryu}}},
  \bibinfo{journal}{\apj} \textbf{\bibinfo{volume}{974}}, \bibinfo{eid}{242}
  (\bibinfo{year}{2024}), \eprint{2305.18538}.

\bibitem[{\citenamefont{{Combi} et~al.}(2022)\citenamefont{{Combi}, {Lopez
  Armengol}, {Campanelli}, {Noble}, {Avara}, {Krolik}, and
  {Bowen}}}]{Combi2022}
\bibinfo{author}{\bibfnamefont{L.}~\bibnamefont{{Combi}}},
  \bibinfo{author}{\bibfnamefont{F.~G.} \bibnamefont{{Lopez Armengol}}},
  \bibinfo{author}{\bibfnamefont{M.}~\bibnamefont{{Campanelli}}},
  \bibinfo{author}{\bibfnamefont{S.~C.} \bibnamefont{{Noble}}},
  \bibinfo{author}{\bibfnamefont{M.}~\bibnamefont{{Avara}}},
  \bibinfo{author}{\bibfnamefont{J.~H.} \bibnamefont{{Krolik}}},
  \bibnamefont{and} \bibinfo{author}{\bibfnamefont{D.}~\bibnamefont{{Bowen}}},
  \bibinfo{journal}{\apj} \textbf{\bibinfo{volume}{928}}, \bibinfo{eid}{187}
  (\bibinfo{year}{2022}), \eprint{2109.01307}.

\bibitem[{\citenamefont{{Guti{\'e}rrez}
  et~al.}(2022)\citenamefont{{Guti{\'e}rrez}, {Combi}, {Noble}, {Campanelli},
  {Krolik}, {L{\'o}pez Armengol}, and {Garc{\'\i}a}}}]{Gutierrez2022}
\bibinfo{author}{\bibfnamefont{E.~M.} \bibnamefont{{Guti{\'e}rrez}}},
  \bibinfo{author}{\bibfnamefont{L.}~\bibnamefont{{Combi}}},
  \bibinfo{author}{\bibfnamefont{S.~C.} \bibnamefont{{Noble}}},
  \bibinfo{author}{\bibfnamefont{M.}~\bibnamefont{{Campanelli}}},
  \bibinfo{author}{\bibfnamefont{J.~H.} \bibnamefont{{Krolik}}},
  \bibinfo{author}{\bibfnamefont{F.}~\bibnamefont{{L{\'o}pez Armengol}}},
  \bibnamefont{and}
  \bibinfo{author}{\bibfnamefont{F.}~\bibnamefont{{Garc{\'\i}a}}},
  \bibinfo{journal}{\apj} \textbf{\bibinfo{volume}{928}}, \bibinfo{eid}{137}
  (\bibinfo{year}{2022}), \eprint{2112.09773}.

\bibitem[{\citenamefont{Shapiro}(2013)}]{shapiro_accretion_2013}
\bibinfo{author}{\bibfnamefont{S.~L.} \bibnamefont{Shapiro}},
  \bibinfo{journal}{Physical Review D} \textbf{\bibinfo{volume}{87}},
  \bibinfo{pages}{103009} (\bibinfo{year}{2013}), ISSN
  \bibinfo{issn}{1550-7998, 1550-2368},
  \urlprefix\url{https://link.aps.org/doi/10.1103/PhysRevD.87.103009}.

\bibitem[{\citenamefont{Peng et~al.}(2024)\citenamefont{Peng, Franchini,
  Bonetti, Sesana, and Chen}}]{Peng:2024wqf}
\bibinfo{author}{\bibfnamefont{P.}~\bibnamefont{Peng}},
  \bibinfo{author}{\bibfnamefont{A.}~\bibnamefont{Franchini}},
  \bibinfo{author}{\bibfnamefont{M.}~\bibnamefont{Bonetti}},
  \bibinfo{author}{\bibfnamefont{A.}~\bibnamefont{Sesana}}, \bibnamefont{and}
  \bibinfo{author}{\bibfnamefont{X.}~\bibnamefont{Chen}}
  (\bibinfo{year}{2024}), \eprint{2411.16070}.

\bibitem[{\citenamefont{{Peng} and {Chen}}(2023)}]{PengChen2023}
\bibinfo{author}{\bibfnamefont{P.}~\bibnamefont{{Peng}}} \bibnamefont{and}
  \bibinfo{author}{\bibfnamefont{X.}~\bibnamefont{{Chen}}},
  \bibinfo{journal}{\apj} \textbf{\bibinfo{volume}{950}}, \bibinfo{eid}{3}
  (\bibinfo{year}{2023}), \eprint{2211.05798}.

\bibitem[{\citenamefont{{Clyburn} and {Zrake}}(2024)}]{Clyburn2024}
\bibinfo{author}{\bibfnamefont{M.}~\bibnamefont{{Clyburn}}} \bibnamefont{and}
  \bibinfo{author}{\bibfnamefont{J.}~\bibnamefont{{Zrake}}},
  \bibinfo{journal}{arXiv e-prints} \bibinfo{eid}{arXiv:2405.10281}
  (\bibinfo{year}{2024}), \eprint{2405.10281}.

\bibitem[{\citenamefont{{S{\'a}nchez-Salcedo}}(2020)}]{Sanchez2020}
\bibinfo{author}{\bibfnamefont{F.~J.} \bibnamefont{{S{\'a}nchez-Salcedo}}},
  \bibinfo{journal}{\apj} \textbf{\bibinfo{volume}{897}}, \bibinfo{eid}{142}
  (\bibinfo{year}{2020}), \eprint{2006.10206}.

\end{thebibliography}
\appendix

\section{Full Posteriors}\label{app:posteriors}
We performed a full Bayesian parameter estimation in the three studied cases, using the vacuum parameters $(\log m_1,\log m_2,a,p_0,\Phi_{\phi\,0})$, sky position and spin inclination $(D_L,\,\cos\theta_K,\,\Phi_K,\,\cos\theta_S,\,\Phi_S)$ and power law parameters $(A,n)$.
Here we report the full posteriors for the three studied cases.
\begin{figure*} 
    \centering    \includegraphics[width=\textwidth]{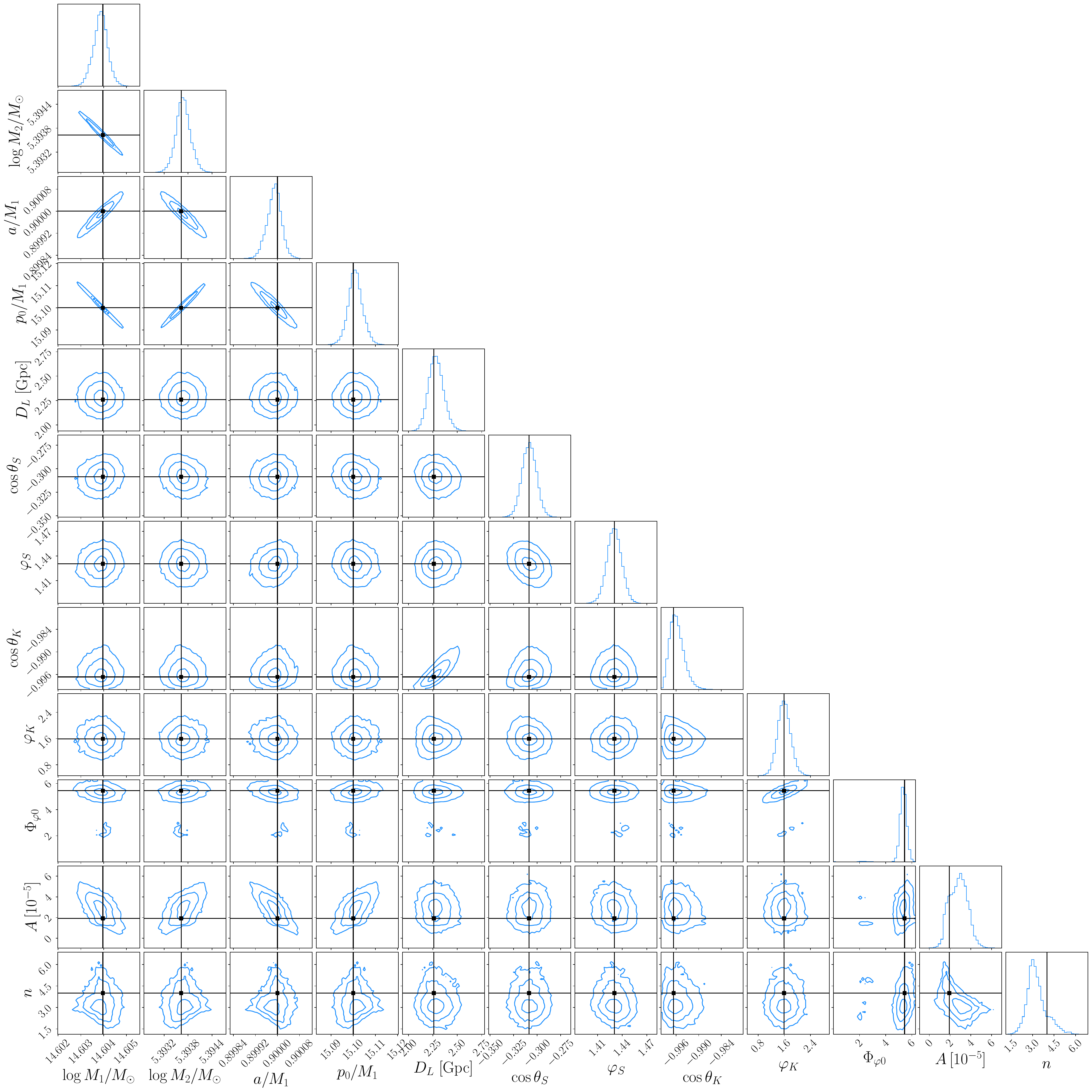}
    \caption{ Full posterior for the case 1 injection. In black we report the injection values of the binary parameters and the expected power law amplitude and slope.
   The power law parameters are estimated by averaging the numerical simulations~\cite{derdzinskiEvolutionGasDisc2021}.
   }
    \label{fig:posterior_q1e4}
\end{figure*}

\begin{figure*}
    \centering
    \includegraphics[width=\textwidth]{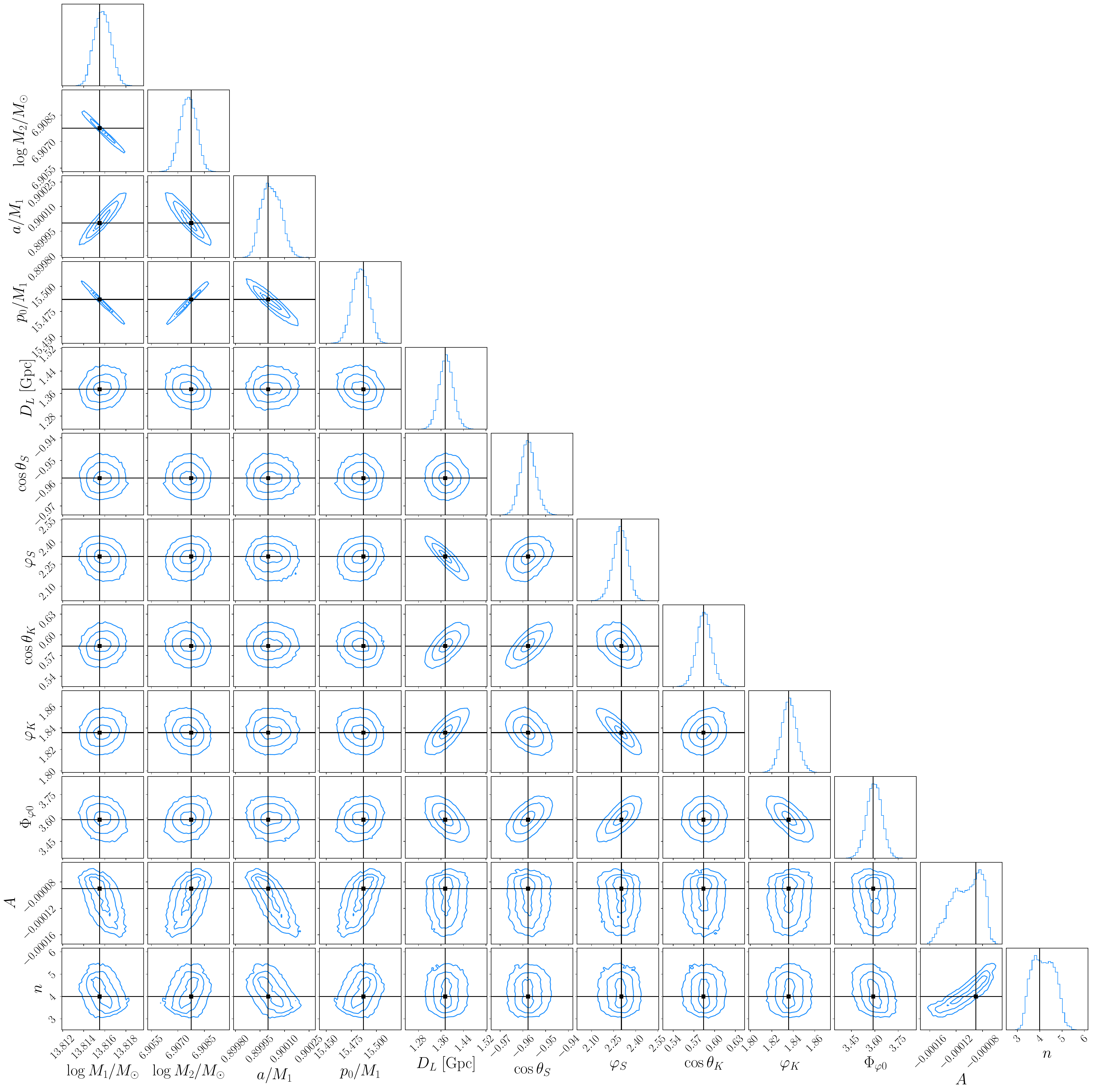}    
    \caption{Full posterior for the case 2 injection. In black we report the injection values of the binary parameters and the expected power law amplitude and slope.
   The power law parameters are estimated by averaging the numerical simulations~\cite{derdzinskiEvolutionGasDisc2021}.}
    \label{fig:posterior_q1e3_200}
\end{figure*}

\begin{figure*}
    \centering
    \includegraphics[width = \textwidth]{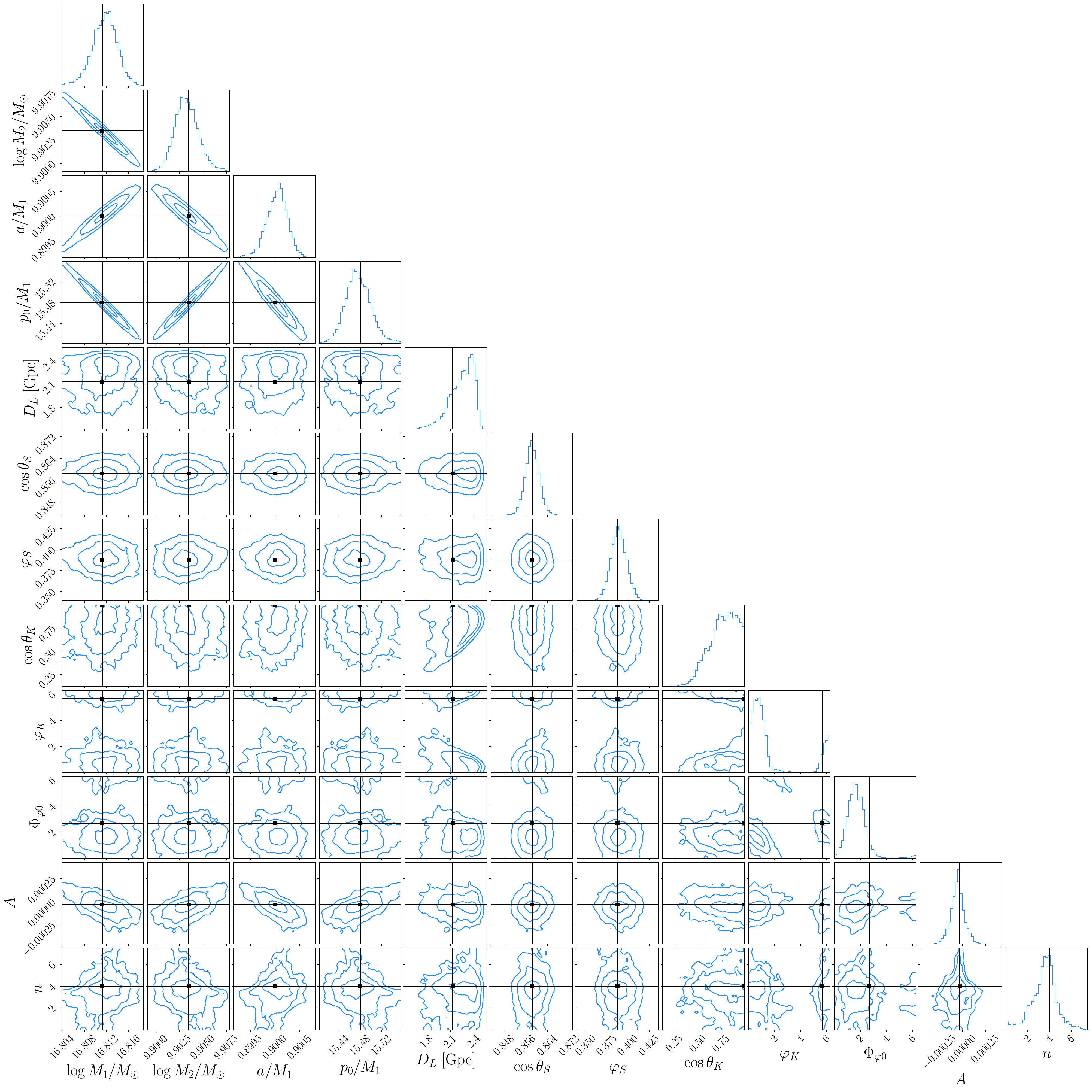}
    \caption{Full posterior for the case 3 injection. In black we report the injection values of the binary parameters and the expected power law amplitude and slope.
   The power law parameters are estimated by averaging the numerical simulations~\cite{derdzinskiEvolutionGasDisc2021}.
    Although the vacuum parameters are well reconstructed in this case, the stochastic features in the environmental torque wash away the power-law contribution.
    }
    \label{fig:posterior_q1e3_a01}
\end{figure*}

\end{document}